\documentclass[fleqn,usenatbib]{mnras}

\usepackage{newtxtext,newtxmath}

\usepackage[T1]{fontenc}

\DeclareRobustCommand{\VAN}[3]{#2}
\let\VANthebibliography\thebibliography
\def\thebibliography{\DeclareRobustCommand{\VAN}[3]{##3}\VANthebibliography}

\usepackage{graphicx}
\usepackage{amsmath}
\usepackage{calligra}
\usepackage{siunitx}

\newcommand{\be}{\begin{equation}}
\newcommand{\ee}{\end{equation}}

\def\ba{\begin{eqnarray}}
\def\ea{\end{eqnarray}}

\def\msol{M_\odot}

\def\he3{^3He}

\def\ltsima{$\; \buildrel < \over \sim \;$}
\def\simlt{\lower.5ex\hbox{\ltsima}}
\def\gtsima{$\; \buildrel > \over \sim \;$}
\def\simgt{\lower.5ex\hbox{\gtsima}}
\def\vel{{\rm v}}
\def\velvec{{\rm \mathbf v}}

\newcommand{\Rstar}{R_{\rm star}}
\newcommand{\Vrms}{\vel_\mathrm{rms}}
\newcommand{\AvgS}[1]{\left< {#1} \right>_{\mathcal{S}}}

\newcommand{\AvgT}[1]{\left< {#1} \right>_t}

\newcommand{\uHz}[1]{\SI{#1}{\micro\hertz}}

\def\Fsingle{F_{\rm wave}}

 \renewcommand{\vec}[1]{\mathbf{#1}}

\title[Two-dimensional simulations of IGWs in a 5 $M_{\odot}$ model]{Two-dimensional simulations of internal gravity waves in a 5 $M_{\odot}$ Zero-Age-Main-Sequence model}

\author[A. Le Saux et al.]{
A.~Le~Saux $^{1,2}$\thanks{E-mail: al598@exeter.ac.uk}, I.~Baraffe $^{1,2}$, T.~Guillet$^{1}$, D.~G.~Vlaykov$^{1}$, A. Morison$^{1}$, J.~Pratt$^{3}$, T.~Constantino$^{1}$, T.~Goffrey$^{4}$
\\
$^{1}$University of Exeter, Physics and Astronomy, EX4 4QL Exeter, UK\\
$^{2}$\'Ecole Normale Sup\'erieure, Lyon, CRAL (UMR CNRS 5574), Universit\'e de Lyon, France\\
$^{3}$Lawrence Livermore National Laboratory, 7000 East Ave, Livermore, CA 94550, USA\\
$^{4}$Centre for Fusion, Space and Astrophysics, Department of Physics, University of Warwick, Coventry, CV4 7AL, UK
}

\date{Accepted XXX. Received YYY; in original form ZZZ}

\pubyear{2022}

\begin{document}
\label{firstpage}
\pagerange{\pageref{firstpage}--\pageref{lastpage}}
\maketitle

\begin{abstract}
Main-sequence intermediate-mass stars present a radiative envelope that supports internal gravity waves (IGWs). Excited at the boundary with the convective core, IGWs propagate towards the stellar surface and are suspected to impact physical processes such as rotation and chemical mixing. Using the fully compressible time-implicit code MUSIC, we study IGWs in two-dimensional simulations of a zero-age-main-sequence 5 solar mass star model up to 91\% of the stellar radius with different luminosity and radiative diffusivity enhancements. Our results show that low frequency waves excited by core convection are strongly impacted by radiative effects as they propagate. This impact depends on the radial profile of radiative diffusivity which increases by almost 5 orders of magnitude between the centre of the star and the top of the simulation domain. In the upper layers of the simulation domain, we observe an increase of the temperature. Our study suggests that this is due to heat added in these layers by IGWs damped by radiative diffusion. We show that non-linear effects linked to large amplitude IGWs may be relevant just above the convective core. Both these effects are intensified by the artificial enhancement of the luminosity and radiative diffusivity, with enhancement factors up to $10^4$ times the realistic values. Our results also highlight that direct comparison between numerical simulations with enhanced luminosity and observations must be made with caution. Finally, our work suggests that thermal effects linked to the damping of IGWs could have a non-negligible impact on stellar structure.
\end{abstract}

\begin{keywords}
stars: oscillations -- software: simulations -- hydrodynamics -- waves -- stars:interior -- asteroseismology
\end{keywords}



\section{Introduction}

Studying the properties of internal gravity waves (IGWs) in stars and planets is of crucial importance as they are supposed to be involved in many physical processes linked to rotation, mixing and magnetism. In the Earth atmosphere, it is generally accepted that they are particularly efficient at transporting angular momentum (AM), energy and chemical elements \citep[see][for a review]{Sutherland2010}. Therefore, stellar physicists have considered their existence in stars since the 1960s \citep{Stein1967} and studied their properties and how they could be excited in this context. \citet{Stein1967} extended the pioneering work on acoustic waves by \citet{Lighthill1952} to a stratified medium, and concluded that turbulent convection can generate internal gravity waves that will propagate in an adjacent stably stratified region. In this work, we consider IGWs excited by this mechanism.

Since then, IGWs have been considered to be involved in various phenomena in stellar interiors. For instance, they have been suggested as a possible explanation for the solid rotation of the solar core inferred from helioseismology \citep[see for example][]{Charbonnel2005}, the mixing profile in the radiative zones of intermediate-mass stars induced from asteroseismology \citep{Pedersen2021} or the slowing down of the core rotation of subgiant stars \citep{Pincon2017} as observed by the \textit{Kepler} mission \citep{Borucki2010}. 

However, the properties of IGWs excited by turbulent convection in stellar interiors remain poorly constrained due to the challenge of their observations. There have been claims of detection in the Sun \citep[see for example][]{Appourchaux2010} and in more massive stars \citep[see for example][]{Aerts2015}, but these are still a matter of debate. Observations of these waves are essential to probe the deeper layers of stars. In the Sun, they will allow the measurement of the rotation profile of the inner 20\% of the solar radius that remain inaccessible to acoustic modes \citep{Garcia2007}. In more massive stars they would bring new constraints on internal rotation, mixing and convective boundary mixing, which has been a long-standing puzzle of stellar structure \citep[see for example][]{Renzini1987}. These more massive stars with a convective core are probably better targets for observations of IGWs. 

Unlike main sequence low-mass stars where the radiative zone is located in the inner part of the star, more massive main sequence stars with $M \gtrsim 2 M_{\odot}$ present a radiative envelope. Internal gravity waves generated at the edge of the convective core propagate towards the surface through a medium of decreasing density, which tend to increase their amplitude. These waves are also damped by radiative diffusion as they travel, and are suggested to transport energy and angular momentum and mix chemical elements through this mechanism \citep{Schatzman1993}. Thus, the evolution of IGWs amplitude will depend on the interplay between growth due to decreasing density and decay due to radiative damping. It remains unclear if these waves should be able to propagate up to the surface. 
More than a decade ago, \citet{Blomme2011} observed a low frequency power excess in the spectra of O type stars observed by the CoRoT \citep{Auvergne2009} mission. They concluded that the physical origin of this power excess was unclear. In recent studies, \citep{Bowman2019, Bowman2020} claim that this low frequency power excess in the spectrum of O and B type stars observed by the CoRoT and TESS \citep{Ricker2015} satellites is due to IGWs excited by turbulent core convection. This hypothesis is supported by hydrodynamical simulations in 3D \citep{Edelmann2019} and 2D \citep{Ratnasingam2020}. However, theoretical work by \citet{Lecoanet2019} and numerical simulations by \citet{Lecoanet2021} do not agree with this conclusion and state that the origin of this power excess is more probably due to a near surface convection zone \citep{Cantiello2021}. This question will remain difficult to answer while the properties of IGWs in such stars remain poorly known.

Hydrodynamical simulations offer a great opportunity to test theoretical models and guide observations. Particularly numerical modelling of internal waves, is a good way to get constraints on the spectrum generated by convection, their amplitude and damping rate within the star.  This has already proven to be efficient in simulations of solar-like stars \citep[see for example][]{Rogers2005, Alvan2014} as well as stars with convective cores \citep[see for example][]{Rogers2013, Horst2020}.
In this work, we present two-dimensional simulations of a 5 solar mass star with different enhancement factors for the luminosity and the radiative diffusivity, as it is widely used for hydrodynamical simulations. This is usually referred to as \textit{boosting} a numerical model. In this study we focus mainly on the damping of IGWs and the impact of boosting on these waves, and we neglect the effect of rotation, even though most OB stars may be moderate to fast rotators. In Sect. \ref{sec:num_sim} we present the numerical set-up of our simulations. Then Sect. \ref{sec:velocities} and \ref{sec:waves_energy_flux} focus on amplitude related properties of IGWs, whereas Sect \ref{sec:non-lin} studies the non-linearity of waves and Sect. \ref{sec:wave_heating} an increase of the temperature observed in the upper layers. Finally, in Sect. \ref{sec:discussion} we summarize and discuss our results.


\section{Numerical simulations}
\label{sec:num_sim}

We performed two-dimensional simulations of the interior of a 5$M_{\odot}$ star model at zero-age-main-sequence (ZAMS) with the fully compressible time-implicit code MUSIC \citep{Viallet11, Viallet13, Viallet16, Geroux16, Goffrey17, Pratt2016, Pratt2017, Pratt2020}. The code solves the inviscid hydrodynamical equations in a fully compressible medium

\begin{equation}
\frac{\partial \rho}{\partial t} = - \vec{\nabla} \cdot (\rho \vec \velvec),
\end{equation}

\begin{equation}
\frac{\partial \rho e}{\partial t} = - \vec{\nabla} \cdot (\rho e \vec\velvec) - p \vec{\nabla} \cdot \vec \velvec - \vec{\nabla} \cdot \vec{F_r} + Q_{\rm nuc},
\end{equation}

\begin{equation}
\frac{\partial\rho \vec\velvec}{\partial t} = - \nabla \cdot (\rho \vec \velvec \otimes \vec \velvec) - \vec{\nabla}p + \rho\vec{g},
\end{equation}
where $\rho$ is the density, $e$ the specific internal energy, $\vec \velvec$ the velocity field, $p$ the gas pressure, $Q_{\rm nuc}$ the nuclear energy rate and $\vec{g}$ the gravitational acceleration, which we assume is radial. For the stellar simulations considered in this work, the major heat transport that contributes to thermal conductivity  is radiative transfer characterised by the radiative flux  $\vec{F_r}$, given within the diffusion approximation by

\begin{equation}
\vec{F_r} = - \frac{4acT^3}{3\kappa \rho} \vec{\nabla} T = - \chi \vec{\nabla} T,
\label{eq:radiative_flux}
\end{equation}
with $\kappa$ is the Rosseland mean opacity of the gas and $\chi$  the radiative conductivity. Realistic stellar opacities and equation of states appropriate for the description of stellar interiors are implemented in MUSIC. Opacities are interpolated from the OPAL tables \citep{Iglesias1996} for solar metallicity, and the equation of state is based on the OPAL EOS tables of \citet{Rogers2002}.

\subsection{Initial stellar model}
The initial structures for the two-dimensional simulations come from a 5 $M_{\odot}$ stellar model computed with the one-dimensional Lyon stellar evolution code \citep{Baraffe1991, Baraffe1998}, using the same opacities and equation of state as implemented in MUSIC. The initial 1D model has an initial helium abundance in mass fraction Y=0.28 and solar metallicity Z=0.02 and is evolved through the pre-main-sequence (PMS) and the early main-sequence. The initial structure of the simulations have burnt 1\% of their hydrogen since the zero-age-main-sequence. There is no overshooting or diffusion considered during the computation of the 1D model. In the MUSIC simulations, the energy generated by nuclear reactions are taken into account in the energy equation as a source term. The nuclear energy profile is the one from the 1D model, and it is assumed to remain constant during the run of the MUSIC simulations.
The properties of the initial 1D model are summarised in Table \ref{tab0}.

\begin{table}
   \caption{Properties of the initial stellar model used for the 2D hydrodynamical simulations.}
   \label{tab0}
   \centering
   \begin{tabular}{c c c c c } 
     \hline \hline
     $M/\msol$ &  $L_{\rm star}/L_\odot^{(a)}$ & $\Rstar$ (cm) &  $r_{\rm conv}/R_{\rm tot}$ & $H_{P,{\rm conv}}$ (cm) \\
      \hline
      5 &  523 & 1.8424 $\times 10^{11}$ & 0.1814 & 1.828 $\times 10^{10}$ \\
      \hline
      \multicolumn{5}{l}{With mass $M$, luminosity $L_{\rm star}$, radius $\Rstar$, size of the convective} \\ 
      \multicolumn{5}{l}{core $r_{\rm conv}$ and pressure scale height at the convective boundary $H_{P,{\rm conv}}$.}\\
      \multicolumn{3}{l}{$^a$ We use $L_\odot = 3.839 \times 10^{33}$ erg/s.}
   \end{tabular}
\end{table}

\subsection{Spherical-shell geometry and boundary conditions}
A detailed description of the set-up of the simulations can be found in \citet{Baraffe2023}. In this section we only summarize some characteristics relevant to this work.
Two-dimensional simulations are performed in a spherical shell using spherical coordinates, namely $r$ the radius and the polar angle $\theta$, and assuming azimuthal symmetry in the $\phi$-direction. For all simulations, the inner radius $r_{\rm in}$ is set at 0.02 $\Rstar$ and the outer radius $r_{\rm out}$ at 0.91 $\Rstar$. Extension of the numerical domain to the photosphere ($r=\Rstar$) is an open challenge for stellar hydrodynamical simulations given the sharp decrease of the pressure scale height with increasing radius. The domain in the co-latitudinal direction ranges from 0 to $\pi$. We use a uniform grid resolution of $N_r \times N_{\theta}$ = 1322 $\times$ 668 cells. This provides a good resolution of the pressure scale height at the convective boundary $H_{P,\rm conv}/\Delta r \sim 147$, with $\Delta r = 1.242 \times 10^8$ cm, the size of a radial grid cell.

Concerning the radial boundary conditions, we impose a constant radial derivative on the density on the inner and outer radial boundaries, as discussed in \citet{Pratt2016}. For the velocity, we impose reflective conditions at the radial boundaries. For our reference model \textit{ref}, the energy flux at the inner and outer radial boundaries are set to the value of the energy flux at that radius in the initial one-dimensional model. In this work we analyse the impact of enhancing the luminosity and thermal diffusivity by factors of 10, $10^2$, and $10^4$. As in \citet{Baraffe2021} and \citet{LeSaux2022}, for the artificially boosted simulations, the energy flux, and equivalently the luminosity, at the boundaries is multiplied by an enhancement factor, and the Rosseland mean opacities $\kappa$ in MUSIC are decreased by the same factor. At the boundaries in the $\theta$-direction we use reflective boundary conditions for the density, velocity and energy.

The characteristics of the four numerical models used in this work are presented in Table \ref{tab1}. We define the convective turnover time $\tau_{\rm conv}$ by
\begin{equation}
\tau_{\rm conv} \coloneqq  \left< \int_{r_{\rm in}}^{r_{\rm conv}} \frac{{\rm d} r} { \Vrms(r,t)  } \right>_t   = \omega_{\rm conv}^{-1},
\label{eq:tau}
\end{equation}
where $r_{\rm in} = 0.02 \Rstar$ is the inner boundary of the two-dimensional simulations and $r_{\rm conv}= 0.1814 \Rstar$ is the location of the convective boundary of the stellar model as defined by the Schwarzschild criterion. The root-mean-square of the velocity $\Vrms$ is defined as
\begin{equation}
    \Vrms(r, t) \coloneqq \sqrt{\AvgS{\velvec^2(r,\theta, t)}},
    \label{eq:vrms}
\end{equation}
where $\velvec$ is the velocity vector. The operators $\AvgT{.}$ and $\AvgS{.}$ are time and angular averages. The temporal average is defined as
\begin{equation}
    \left< f(t) \right>_t \coloneqq \frac{1}{T} \int_0^T f(t) {\rm d} t, 
    \label{eq:time_av}
\end{equation}
with $T$ the time of integration considered for the average. The lower limit of the integral $t=0$ corresponds to the time from which convection is in steady state \citep[see][for details on the definition of this steady state.]{Baraffe2023}. The angular average is defined as
\begin{equation}
    \left< g(\theta) \right>_{\mathcal{S}} \coloneqq \frac{1}{4\pi} \int_\mathcal{S} g(\theta) \, 2\pi \sin \theta {\rm d} \theta.
    \label{eq:angular_av}
\end{equation}

In Eq. \eqref{eq:tau} we also define the convective turnover frequency $\omega_{\rm conv}$ which is the frequency associated with the characteristic timescale $\tau_{\rm conv}$. \citet{Aerts2021} provide observed values of the convective turnover frequency for Slowly-Pulsating B (SPB) stars with mass approximately between 3$M_{\odot}$ and 9$M_{\odot}$. They found typical values of $\omega_{\rm conv}$ in the range 0.2 to 0.5 \uHz{}, which is close to the value of \uHz{0.7} in our reference model. 
We consider the stellar model studied in this paper to be a template of intermediate-mass stars with a convective core, with mass between 3 and 20 $M_{\odot}$. Therefore, our conclusions regarding wave damping and propagation could be applied, at least qualitatively, to other stellar masses.

\begin{table}
   \caption{Summary of the two-dimensional simulations.}
   \label{tab1}
   \centering
   \begin{tabular}{l c c c c} 
     \hline \hline
     Simulation &  $L/L_{\rm star}$ & $\tau_{\rm conv}^{(a)}$ (s) &  $N_{\rm conv}^{(b)}$ &  $\omega_{\rm conv}^{(c)}$ (\uHz{})  \\
      \hline
      ref &  1 & 1.5 $\times 10^6$ & 52 & 0.7 \\
       boost1d1 &  10$^1$ & 6.0 $\times 10^5$ & 35 & 1.7 \\
        boost1d2 &  10$^2$ &   3.0 $\times 10^5$ & 58 & 3.4 \\
        boost1d4 &  10$^4$ & 6.5 $\times 10^4$& 54 & 15.4\\
      \hline
      \multicolumn{5}{l}
{$^a$ Convective turnover time (see Eq. \eqref{eq:tau} for its definition),} \\ 
 \multicolumn{5}{l}{measured from our simulations.} \\
\multicolumn{5}{l}{$^b$ Number of convective turnover times used for this work.}\\
\multicolumn{5}{l}{$^c$ Convective turnover frequency associated with $\tau_{\rm conv}$}\\
   \end{tabular}
\end{table}

\subsection{Stratification and radiative diffusivity}
A medium is stably stratified when it is stable against convection. A particle of fluid moved radially in a stratified medium will oscillate around its equilibrium position with a finite amplitude \citep[see for example][]{Lighthill1978}. Therefore, one of the main characteristic of such medium is its ability to support internal waves, for which the restoring force is buoyancy. The maximum frequency of oscillation of this particle is set by the Brunt-Väisälä, or buoyancy, frequency $N$. Expressed in Hertz, it is defined as
\begin{equation}
    N = \frac{1}{2\pi} \sqrt{g \left( \frac{1}{\Gamma_1} \frac{{\rm d}\ln p}{{\rm d}r} - \frac{{\rm d}\ln \rho}{{\rm d}r} \right) },
    \label{eq:BV_freq}
\end{equation}
where $\Gamma_1$ is the first adiabatic exponent,
\begin{equation}
  \Gamma_1 = \left(\frac{{\rm d}\ln \rho}{{\rm d}\ln p}\right)_{\mathrm{ad}}.
\end{equation}{}
A fluid is stably stratified if $N^2 > 0$ and the convective instability arises if $N^2 \leq 0$.

\begin{figure}
    \centering
    \includegraphics[width=0.5\textwidth]{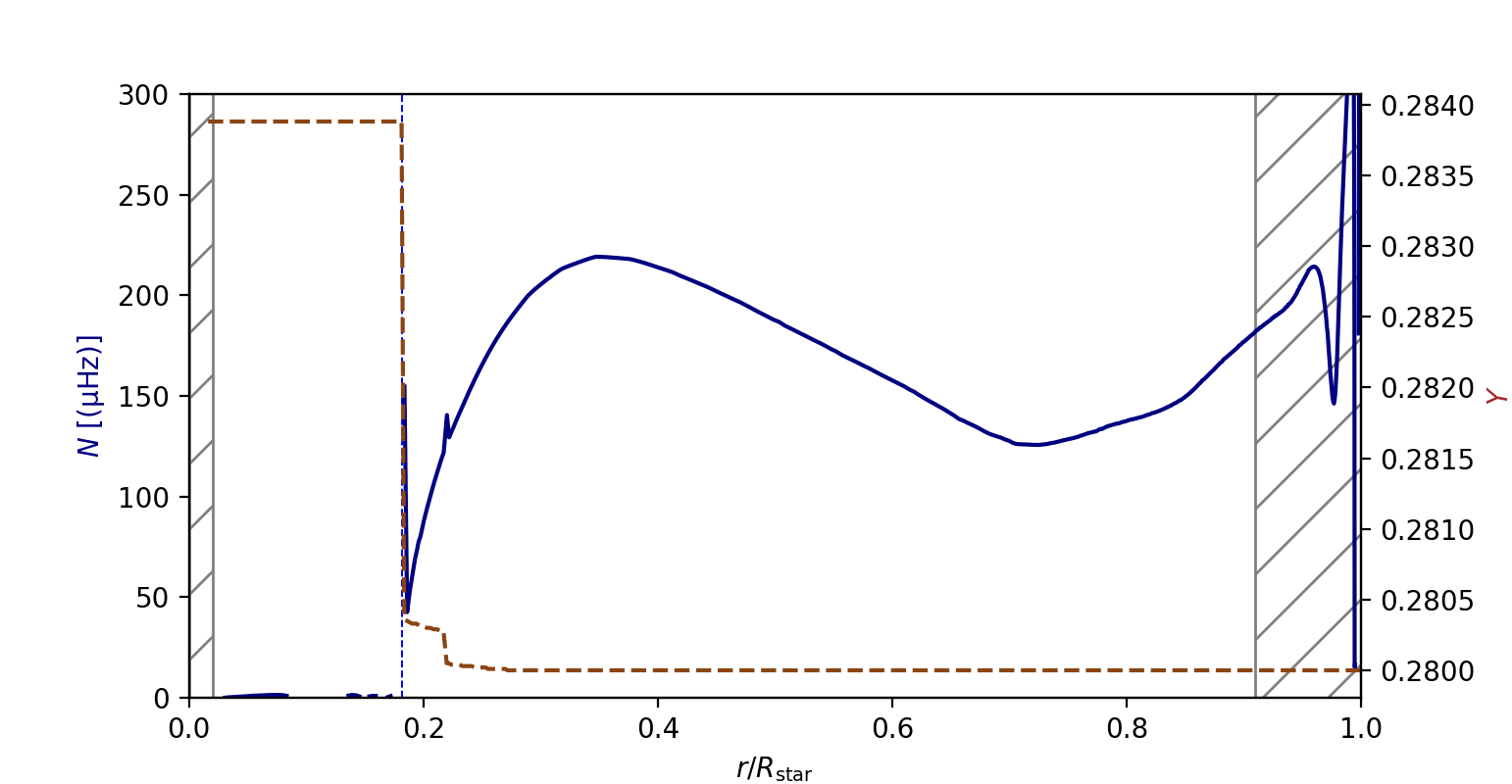}
    \caption{Radial profile of the Brunt-Väisälä frequency (left y-axis, blue curve) and the helium mass fraction $Y$ (right y-axis, brown dashed curve) in the initial 1D model. The hatched regions are not considered in the two-dimensional simulations.}
    \label{fig:BV_profile}
\end{figure}

Figure \ref{fig:BV_profile} shows the radial profile of the Brunt-Väisälä frequency (left y-axis, blue curve) and the helium mass fraction $Y$ (right y-axis, brown dashed curve) in the initial 1D model. The peak of the Brunt-Väisälä frequency just above the convective core is a result of the small gradient of $Y$ in this region. In our simulations the hatched regions are excluded. Internal gravity waves are able to propagate in radiative zones and are evanescent in convective zones. This means they can travel in region where $N^2 > 0$, with the condition on their frequency $\omega < N$. From Fig. \ref{fig:BV_profile} we can see that an IGW of frequency $\omega$ might not propagate in the whole radiative envelope. For instance, a wave with a frequency of $150$ \uHz{} may only propagate between $r\sim 0.25 \Rstar$ and $r\sim 0.6 \Rstar$.

\begin{figure}
    \centering
    \includegraphics[width=0.5\textwidth]{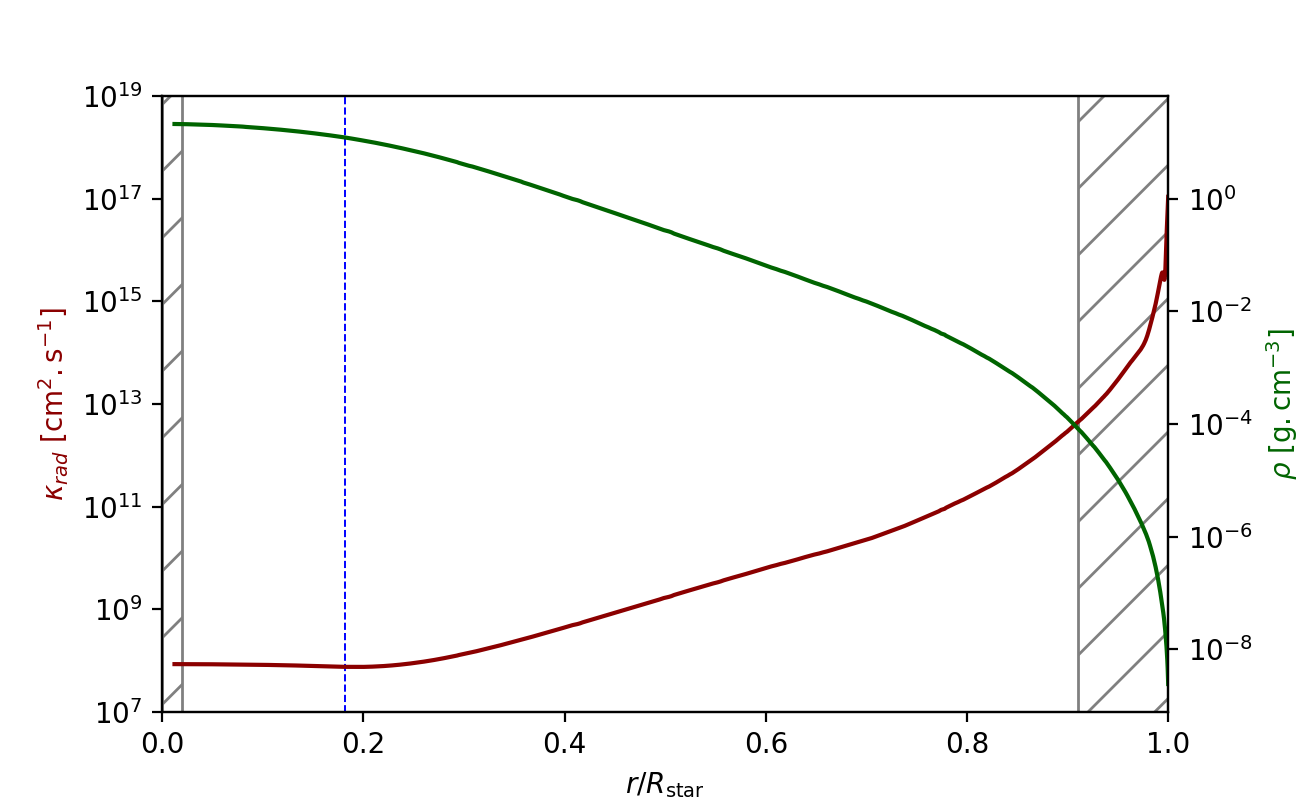}
    \caption{Radial profile of radiative diffusivity $\kappa_{\rm rad}$ (red, left y-axis) and density $\rho$ (green right x-axis) for the initial 1D model. The blue vertical dotted line indicates the convective boundary. The hatched region is not considered in the two-dimensional simulations.}
    \label{fig:Krad_profile}
\end{figure}

The radiative diffusivity is of major importance as it is the main mechanism that damp waves. It is defined as
\begin{equation}
	\kappa_{\rm rad} = \frac{\chi}{\rho c_p},
	\label{eq:Krad}
\end{equation}
with $\chi$ the radiative conductivity defined by Eq.~\eqref{eq:radiative_flux} and $c_p$ the specific heat capacity at constant pressure.
In our reference simulation, model \textit{ref}, we use a realistic profile of radiative diffusivity for a $5 M_{\odot}$ model. This allows modelling of a realistic damping of IGWs. This profile is displayed in Fig. \ref{fig:Krad_profile} (left y-axis). This figure also shows the radial profile of the density in our model (right y-axis). Note that the profiles displayed in Fig. \ref{fig:Krad_profile} are the ones used in our simulation \textit{ref}. This is important as these two quantities define the propagation properties of IGWs.

Indeed, as they propagate towards the surface, IGWs amplitudes grow due to decreasing density and decay due to radiative damping. The variation of both the radiative diffusivity and the density is very large in a star, by 9 orders of magnitude between the centre and the surface. This is one of the main challenges that hydrodynamical simulations have to deal with. In our simulations, that exclude the 9\% outer layers, this variation is restricted to approximately 5 orders of magnitude. These variations have to be taken into account in hydrodynamical simulations because they significantly impact propagation of IGWs as we will see in the next sections.


\section{Velocities}
\label{sec:velocities}
Internal gravity waves manifest themselves as perturbations in density, temperature, luminosity or velocity. In this section, we focus on the radial velocity.

\subsection{Radial velocity pattern}

\begin{figure*}
    \centering
    \includegraphics[width=1.0\textwidth]{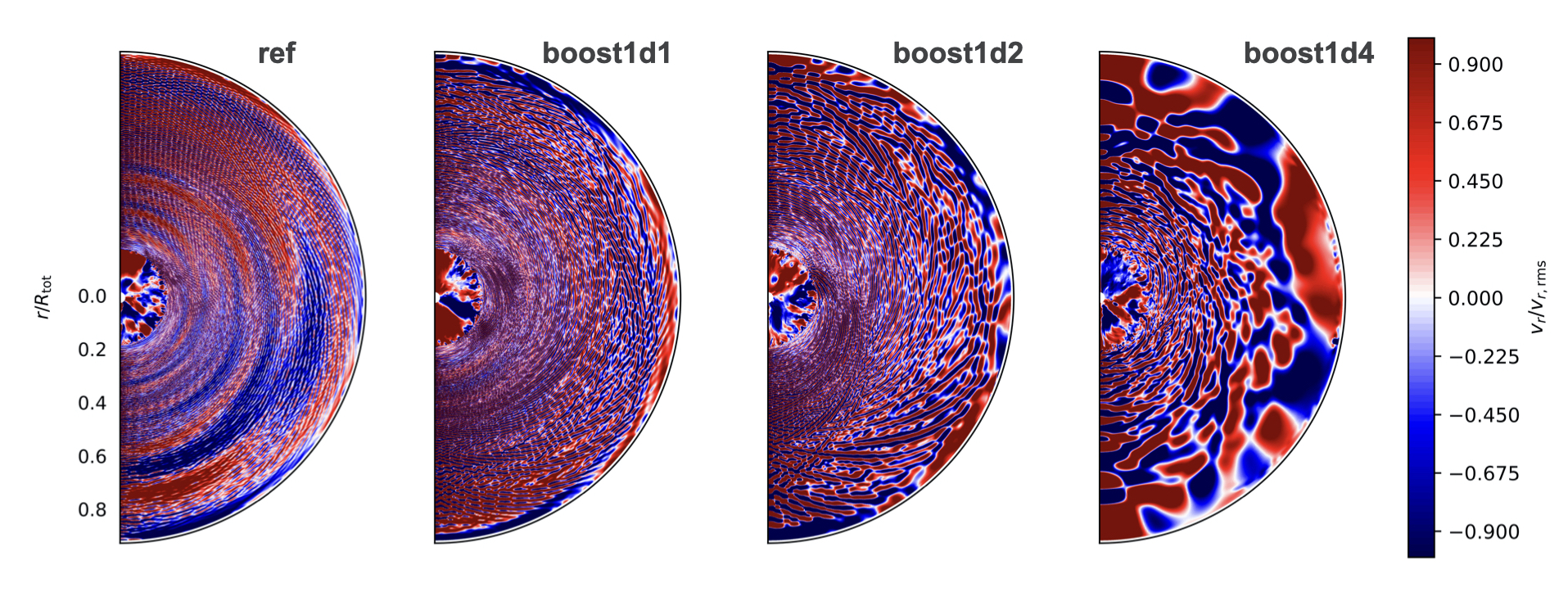}
    \caption{Visualisation of the radial velocity for the four stellar simulations \textit{ref}, \textit{boost1d1}, \textit{boost1d2,} and \textit{boost1d4} as a function of radius, $r$, and co-latitude, $\theta$. The radial velocity is normalised by the rms radial velocity. Positive values of the radial velocity (ref) are outward and negative (blue) are inward.}
    \label{fig:Ur_2d}
\end{figure*}

Figure \ref{fig:Ur_2d} shows snapshots of the radial velocity $\vel_r$ for the four simulations. For better visualisation, the radial velocity is normalised by the root-mean-square value of the radial velocity $\vel_{r, \rm rms}$. The convective core extends from the centre up to $r_{\rm conv}=0.1814 \Rstar$. In this region, the structure of the flows is similar in the four snapshots, with large coherent upflows (red) and downflows (blue).

The three simulations \textit{ref}, \textit{boost1d1} and \textit{boost1d2} present similar patterns in the radiative envelope with the classical spiral structure corresponding to wavefronts of IGWs. The inclination of these wavefronts with respect to the convective boundary, i.e. wrt the horizontal (or angular) direction, defines an angle $\alpha$, to which correspond a particular frequency. This is known as the St. Andrews cross \citep{Sutherland2010} and results from the dispersion relation for IGWs that set the direction of propagation of the waves \citep[see Eq. (7.61) from][]{Vallis2017}
\begin{equation}
    \frac{\omega^2}{N^2} = \frac{k_{\rm h}^2}{k^2} \coloneqq \cos^2\alpha,
    \label{eq:dispersion}
\end{equation}
where $\omega$ is the wave frequency and $k$ the total wavenumber defined by $k = \sqrt{k_r^2 + k_{\rm h}^2}$, with $k_r$ its radial part and $k_{\rm h}$ its horizontal part defined at a given radius $r$ by:
\begin{equation}
    k_{\rm h}^2 \coloneqq \frac{\ell(\ell+1)}{r^2},
    \label{eq:kh}
\end{equation}
with $\ell \geq 0$ the spherical harmonic, or angular, degree. Figure \ref{fig:Ur_2d} suggests that the angle $\alpha$ increases with the boosting factor. Therefore, waves of higher frequencies dominate in the radiative zone when the luminosity is increased, an effect already suggested by \citet{Stein1967} in their analytical work. This was also observed in the solar-like star simulations from \citet{LeSaux2022}.
The structure of the radiative envelope of \textit{boost1d4} does not present the classical spiral IGWs wavefronts pattern. The structure in this most boosted case is larger scale and not periodic any more. Similar patterns are observed in other simulations of intermediate-mass stars using the same or different artificial boosting factors for the luminosity and the radiative diffusivity \citep[see for example][]{Edelmann2019,Horst2020}.

\subsection{Radial evolution of the velocity amplitude}
\label{sec:evol_ampl}

According to linear theory, the evolution of the amplitude of an IGW is expected to depend on the stratification of the supporting medium and on spatial damping due to radiative effects as well as on the frequency and wavelength of the wave. The analytical formula that expresses these dependences is \citep{Press1981, Zahn1997}

\begin{equation}
	\vel_r(r,\ell,\omega) = C \rho^{-1/2} k_{\rm h}^{3/2}\left(\frac{N^2-\omega^2}{\omega^2}\right)^{-1/4} \rm e^{-\tau/2},
	\label{eq:ampl_vr_withC}
\end{equation}
where $\omega$ and $N$ are expressed in hertz and $k_{\rm h}$ is the local horizontal wavenumber, already defined in Eq. \eqref{eq:kh}. The exponential term in Eq. \eqref{eq:ampl_vr_withC} represents the effect of radiative damping, and the parameter $\tau$ is defined as
\begin{equation}
    \tau(r,\ell,\omega) = [\ell(\ell+1)]^{3/2} \int_{r_{\rm e}}^r \kappa_{\rm rad} \frac{N^3}{\omega^4} \left(\frac{N^2}{N^2-\omega^2}\right)^{1/2} \frac{{\rm d} r}{r^3},
    \label{eq:radiative_damping}
\end{equation}
where $r_e$ is the radius at which the waves are excited. It is important to keep in mind that in boosted simulations, the radiative diffusivity is enhanced by the same amount as the luminosity. Consequently, wave damping by radiative diffusion is enhanced in a boosted simulation. In Eq. \eqref{eq:ampl_vr_withC}, $C$ is a constant fixing the amplitude. In this work we chose to fix it such as the analytical velocity amplitude of a wave at $r=r_e$ matches the amplitude of the velocity in the simulations, i.e. $\vel_0(\ell, \omega) = \vel_r(r_e,\ell,\omega)$. Therefore, we obtain
\begin{equation}
	C = \vel_0(\ell,\omega) \rho^{1/2}_0 k_{\rm h,0}^{-3/2}\left(\frac{N_0^2-\omega^2}{\omega^2}\right)^{1/4},
	\label{eq:norm_constant}
\end{equation}
with $k_{\rm h,0} = \sqrt{\ell(\ell+1)}/r_e$, $\rho_0 = \rho(r_e)$, $N_0 = N(r_e)$ and by definition $\tau(r_e,\ell,\omega) = 0$.
Finally, we can write the analytical expression of the radial velocity as
\begin{equation}
	\vel_r(r,\ell,\omega) = \vel_0(\ell, \omega) \left(\frac{\rho}{\rho_0}\right)^{-1/2} \left(\frac{k_{\rm h}}{k_{\rm h,0}}\right)^{3/2} \left(\frac{N^2-\omega^2}{N_0^2-\omega^2}\right)^{-1/4} \rm e^{-\tau/2}.
	\label{eq:ampl_vr}
\end{equation}
This formula is similar to the one obtained by \citet{Ratnasingam2019}.
In our simulations, we determine the amplitude of the radial velocity at a given frequency $\omega$ and angular degree $\ell$ using a temporal Fourier transform and a decomposition on the spherical harmonic basis of the velocities computed by MUSIC. The definition we are using for spherical harmonics and Fourier transform are the same as in \citet{LeSaux2022} (see their Appendix A). We obtain the power spectrum of the radial velocity $P[\hat{\vel}_r^2](r,\ell,\omega)$ which scales as the amplitude squared of a given mode ($\ell$, $\omega$).

\begin{figure}
    \centering
    \includegraphics[width=0.49\textwidth]{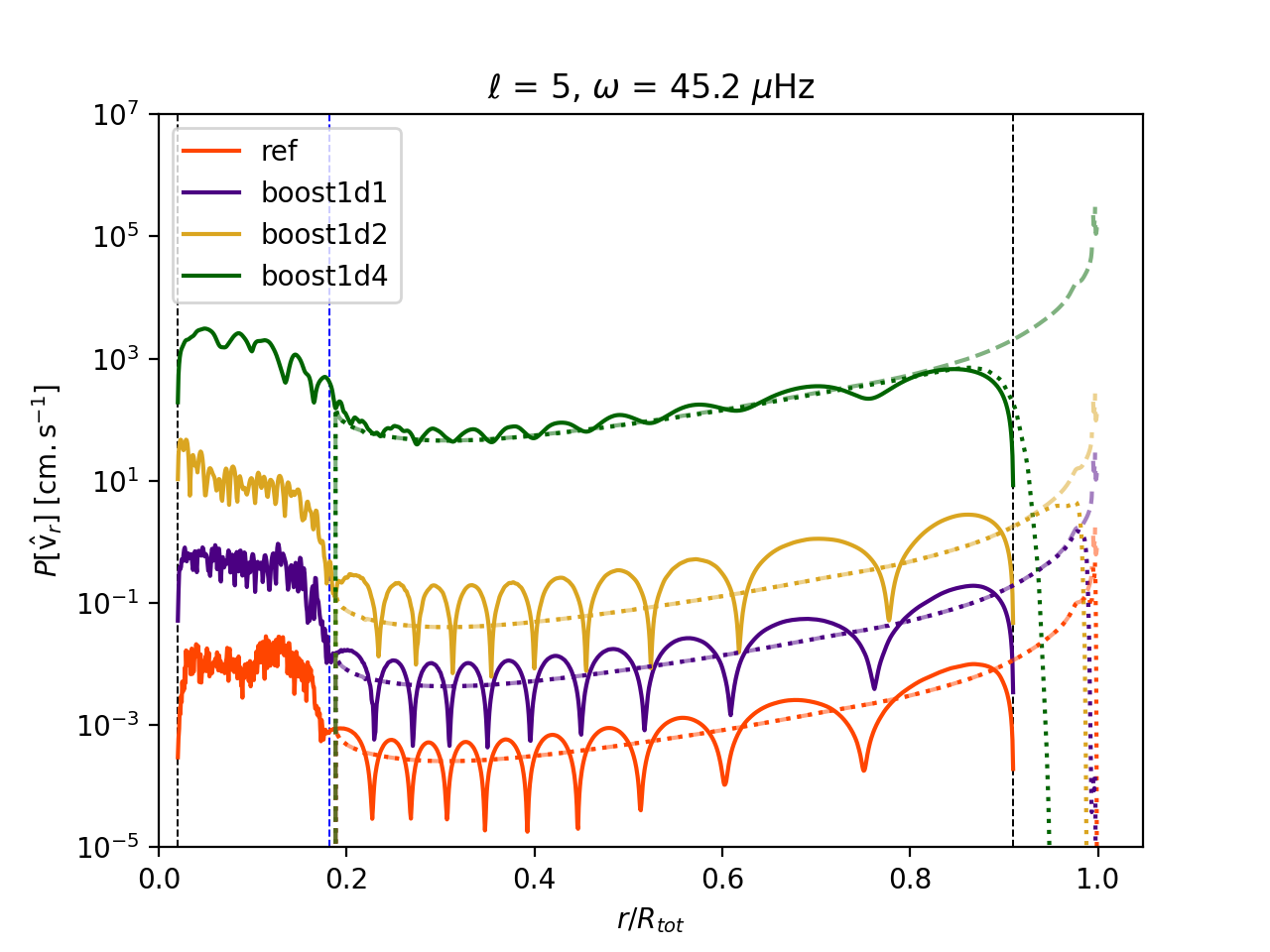}
    \caption{Wave amplitude as a function of normalised radius for the four simulations \textit{ref} (orange), \textit{boost1d1} (indigo), \textit{boost1d2} (yellow), and \textit{boost1d4} (green) for angular degree $\ell = 5$ and frequencies $\omega = 45.2$ \uHz{}. The vertical black lines indicate the boundaries of the simulation domain and the vertical blue line indicate the convective boundary.  The solid lines are the velocity measured in the simulations. Dotted and dashed lines are the theoretical velocity amplitudes computed with Eq. \eqref{eq:ampl_vr} with and without the damping term, respectively.}
    \label{fig:wave_amplitude}
\end{figure}

Figure \ref{fig:wave_amplitude} compares the analytical expression from Eq. \eqref{eq:ampl_vr} (dotted lines) to the corresponding wave velocity amplitude from MUSIC simulations, $\sqrt{P[\hat{\vel}_r^2]}$ (solid lines), as a function of normalised radius for the four numerical models. We have also included the analytical velocity amplitude without the damping term (dashed lines), i.e. we are neglecting the term $\rm e^{-\tau/2}$ in Eq. \eqref{eq:ampl_vr}. The spatial boundaries of the simulation domain at $r_{\rm in} =0.02 \Rstar$ and $r_{\rm out} =0.91 \Rstar$ are specified by the vertical black dashed lines, and the convective boundary $r_{\rm conv}=0.1814 \Rstar$ is indicated by the vertical blue dashed line. For the analytical expressions, we need to set the initial velocity amplitude of the waves in the excitation region, i.e. close to the convective boundary. We arbitrarily chose for $\vel_0$, the value of the radial velocity in the simulations at $r_e = 0.183 \Rstar$, just above the convective core.
Figure \ref{fig:wave_amplitude} presents the velocity amplitude of a wave with angular degree $\ell = 5$ and frequency $\omega = 45.2$ \uHz{}. In the four simulations, the wave amplitude shows a similar oscillatory pattern, where the troughs are the radial nodes of the corresponding standing wave. To form these standing waves, propagating IGWs generated by core convection travel towards the surface of the star until they reflect at their outer turning point or at the top of the simulation domain. A turning point is defined as the radius where $k_r^2 = 0$, or equivalently $\omega = N$ from Eq. \eqref{eq:dispersion}. Travelling back towards the centre they again reflect at their inner turning point. Travelling back and forth, propagating waves of a given frequency and angular degree interfere with themselves and form standing waves. These are called gravity modes, or g modes, and have high amplitudes that can be seen as high narrow peaks in the power spectra of the velocity.
For the mode observed in Fig. \ref{fig:wave_amplitude}, the oscillation code GYRE \citep{Townsend2013, Townsend2018,Goldstein2020} predicts a radial order \footnote{By convention, the radial order $n$ is negative for a g mode and positive for a p mode (standing acoustic waves).} $n = -9$, which is also the number of nodes present in the simulations. This confirms that we see g modes in Fig. \ref{fig:wave_amplitude}. 

Compared with the analytical predictions from Eq. \eqref{eq:ampl_vr} (dotted lines in Fig. \ref{fig:wave_amplitude}), the simulation velocities present a very similar global evolution from the convective boundary up to the top of the simulation domain. Except for the oscillations, but these are not taken into account in the linear analytical expression of the velocity amplitude.
However, there are notable differences between the four numerical models.
Firstly, as expected the velocity amplitude of the waves increases with the enhancement of the luminosity.
Secondly, the analytical amplitudes predict a sharp drop at a given radius close to the surface, which corresponds to the location where the wave is totally damped out, i.e. the location where the wave will deposit most of its energy. This abrupt drop is indeed due to radiative damping as it is not present in the case with no damping (dashed lines). For models \textit{ref}, \textit{boost1d1} and \textit{boost1d2} this abrupt drop is approximately at $r \simeq \Rstar$, but for model \textit{boost1d4} it is located around $r \simeq 0.91\Rstar$.
By looking at Eq. \eqref{eq:ampl_vr} we can see the radiative diffusivity, which is enhanced by the same amount as the luminosity, is included in the expression of $\tau$. As a result, damping of waves by radiative diffusion increases in a boosted model. This is why the sharp drop of the dotted lines is not located at the same radius for the four cases. However, for the waves considered, with $\ell = 5$ and $\omega = 45.2$ \uHz{}, this sharp drop is located at radii $r \geq 0.91 \Rstar$, thus they are able to reach the top of the simulation domain.

\begin{figure*}
\centering
   \includegraphics[width=0.49\textwidth]{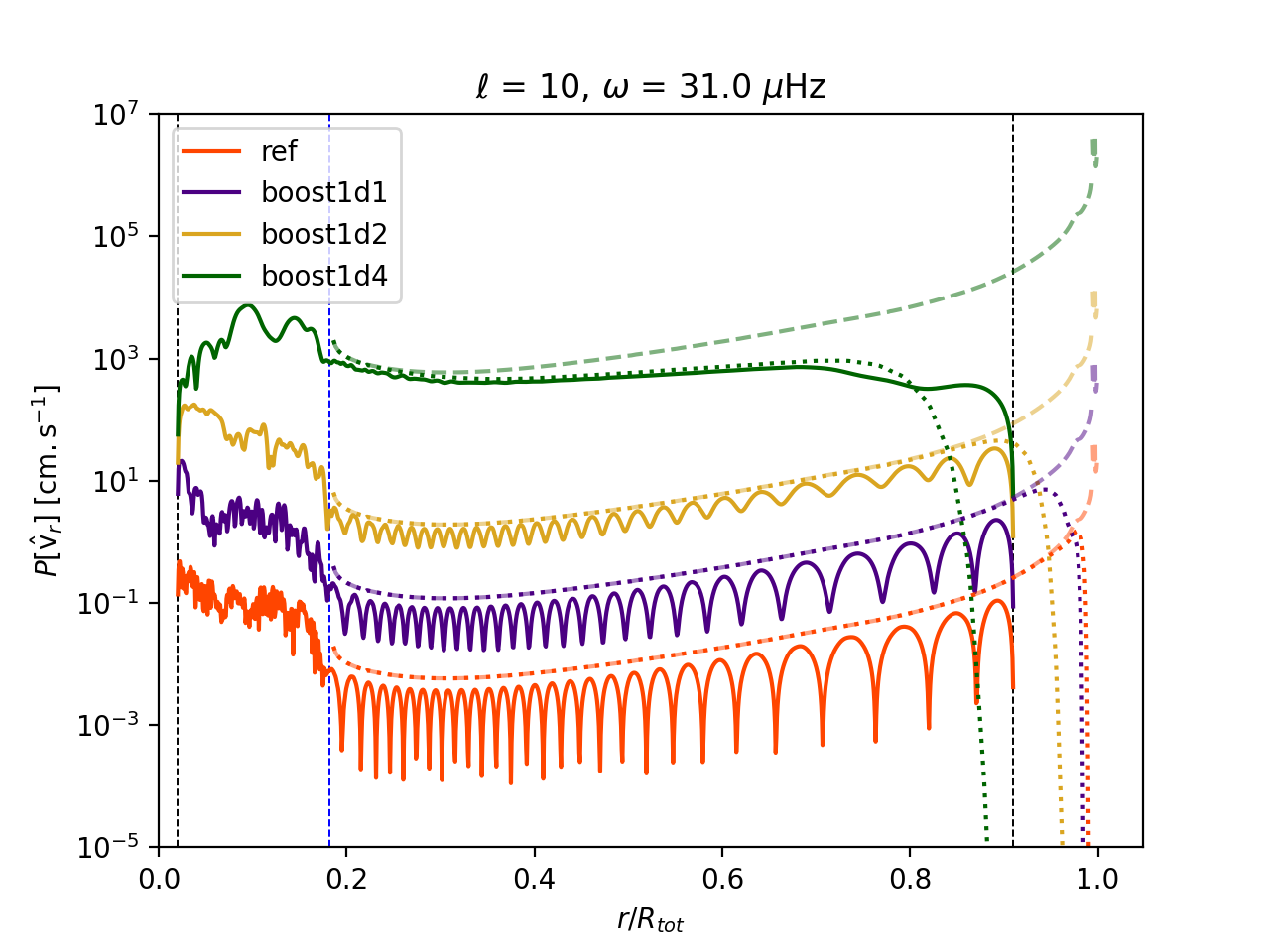}
    \includegraphics[width=0.49\textwidth]{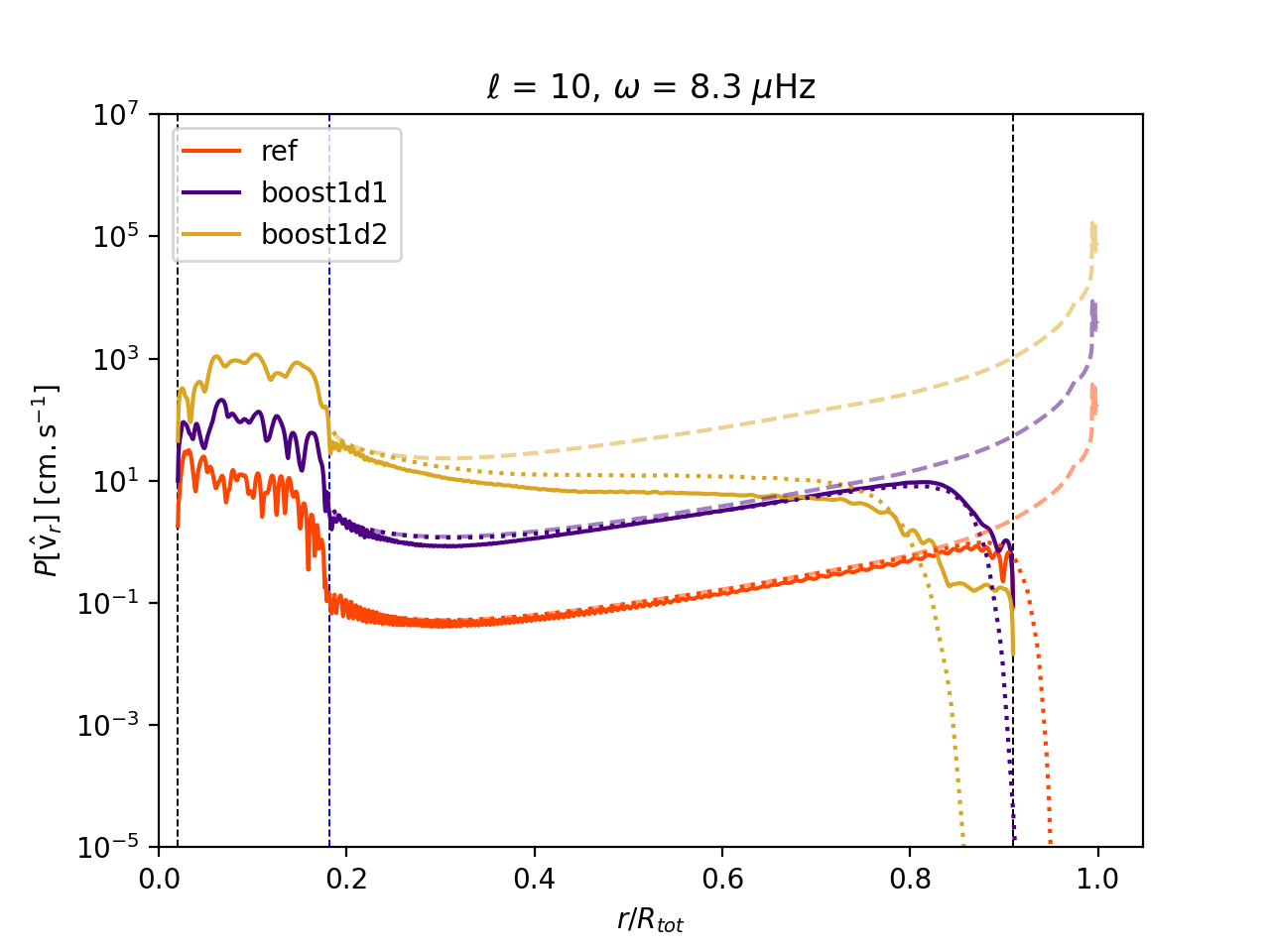}
     \caption{Same as Fig. \ref{fig:wave_amplitude} but for waves with $\ell = 10$ and $\omega = 31.0$ \uHz{} (left panel) and with $\ell = 10$ and $\omega = 8.3$ \uHz{} (right panel)}
     \label{fig:wave_amplitude_comparison}
\end{figure*}

\subsection{Influence of the boost}

From Eqs. \eqref{eq:radiative_damping} and \eqref{eq:ampl_vr}, it is clear that when the angular degree $\ell$ and the frequency $\omega$ are changed, the amplitude and the damping of the corresponding wave are impacted. Indeed, Eq. \eqref{eq:ampl_vr} shows that wave amplitude depends on $\ell$ and $\omega$ and Eq. \eqref{eq:radiative_damping} shows that waves with higher $\ell$ and smaller $\omega$ will be damped more efficiently.
The radial velocity amplitude for a wave with $\ell$ = 10 and $\omega$ = 31.0 \uHz{} is plotted on the left panel of Fig. \ref{fig:wave_amplitude_comparison}. As expected, the damping is more important than for the wave with $\ell$ = 5 and $\omega$ = 45.2 \uHz{} (see Fig. \ref{fig:wave_amplitude}). This is highlighted by the abrupt drop of the theoretical amplitude (dotted lines) which is shifted towards smaller radii in the three boosted models. Indeed, for this wave, the drop is located at $r \simeq 0.95\Rstar$ in model \textit{boost1d1}, $r \simeq 0.91\Rstar$ in model \textit{boost1d2} and $r \simeq 0.8\Rstar$ in model \textit{boost1d4}. However, for model \textit{ref} the location of the drop is not changed, meaning that the effect of radiative damping remains weak for this wave with $\ell$ = 10 and $\omega$ = 31.0 \uHz{}.
Note that for the four simulations the agreement with theory is still relatively good. The case with no damping (dashed lines) is now clearly different from the other two (solid and dotted lines). In that case, the shape of the curves remains the same as in Fig. \ref{fig:wave_amplitude}. Neglecting the damping for these waves would imply that they would be able to propagate up to the surface for all models, yielding an erroneous prediction.
  
On the right panel of Fig. \ref{fig:wave_amplitude_comparison}, the angular degree and frequency are set to $\ell$ = 10 and $\omega$ = 8.3 \uHz{} respectively, implying more efficient damping. Model \textit{boost1d4} is not plotted here as we do not expect such low frequency IGWs to be excited in this simulation. Indeed, the frequency $\omega$ = 8.3 \uHz{} is smaller than the convective frequency, $\omega_{\rm conv}$ = 15.4 \uHz{}, for this simulation (see Table \ref{tab1}). A convective region with associated frequency $\omega_{\rm conv}$ is expected to generate waves with frequencies $\omega \geq \omega_{\rm conv}$ \citep{Lecoanet2013}.
On this plot, the abrupt drop of the analytically predicted velocity amplitude (dotted lines) for models \textit{boost1d1} and \textit{boost1d2} is located at $r \simeq 0.8\Rstar$ and $r \simeq 0.85\Rstar$ respectively. We thus observe a similar phenomenon as for model \textit{boost1d4} in the left panel of Fig. \ref{fig:wave_amplitude_comparison}. Namely, that the waves in the simulations are damped before the top of the numerical domain and do not form g modes.

Our results highlight that the artificial enhancement of the luminosity and the radiative diffusivity of a numerical model impacts not only the amplitude of the waves but also their spatial damping. This is particularly important at low frequencies, as expected from Eq. \eqref{eq:radiative_damping}. This enhanced damping of waves in the low frequency regime in boosted simulations was already suggested by \citet{Horst2020}. When the luminosity is boosted, the increased damping implies that waves over a smaller range of frequencies reach the top of the cavity. Particularly, low frequency g modes are fully damped in boosted simulations compared to the one with realistic luminosity. In addition, waves of given frequency and angular degree will be damped out in different locations in boosted simulations, therefore depositing their energy in different regions. In a case where the luminosity is artificially enhanced but not the radiative diffusivity, as in the simulation of \citet{Horst2020}, the location where IGWs are damped out is not modified. However, because of the higher luminosity, the dominant frequency range of excited waves will be different from in a non boosted case \citep[see][and Sect. \ref{sec:waves_energy_flux}]{LeSaux2022}. Therefore, in simulations with different enhancement factors for the luminosity and radiative diffusivity, we do not expect the propagation of IGWs to be identical as in a model with realistic luminosity.

Finally, comparison between the two analytical cases with and without damping also highlights that running simulations with unrealistic radiative diffusivity may not capture the proper propagation properties of IGWs. Particularly for low frequency waves, which are the ones more impacted by radiative diffusion. In these simulations transport by IGWs should be studied with caution as well as their ability to propagate up to the surface or not.

\subsection{Radial kinetic energy density}
\label{sec:Ekin}

\begin{figure}
    \centering
    \includegraphics[width=0.5\textwidth]{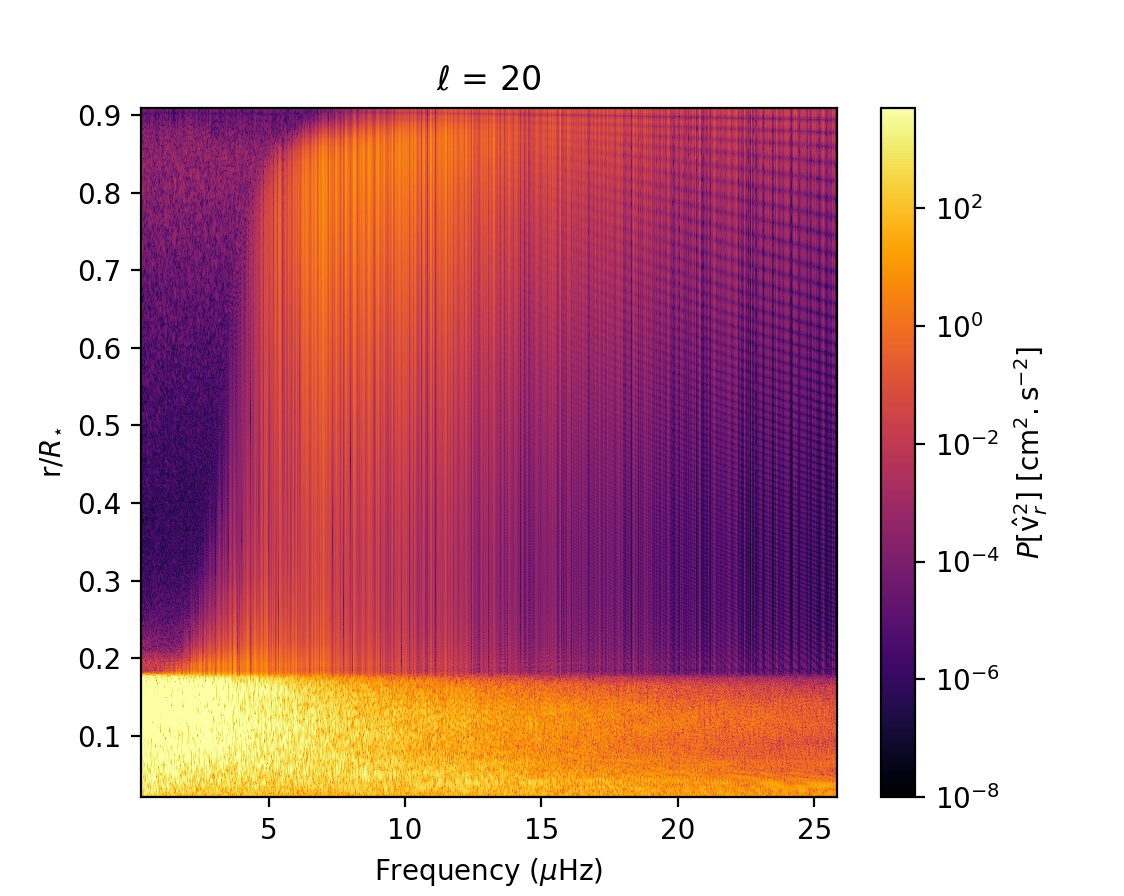}
    \caption{Power spectrum of the radial velocity for the simulation \textit{ref}. The angular degree is fixed at $\ell$= 20. The power spectrum is obtained via mode projection on the spherical harmonics basis and a temporal Fourier transform of the radial velocity.}
    \label{fig:Ekin_ref_ell20}
\end{figure}

Figure \ref{fig:Ekin_ref_ell20} presents the power spectrum of the radial velocity $P[\hat{\vel}_r^2]$, as a function of normalised radius and frequency for an angular degree $\ell$ = 20. This quantity $P[\hat{\vel}_r^2]$ is equivalent to the radial kinetic energy density. This plot offers a general overview on the dependence of radiative damping of IGWs on frequency and radius. In the convective core, between the bottom of the plot and $r_{\rm conv}$, the spectrum is relatively homogeneous at all frequencies and characteristic of a convective zone. The bright ridges observed in the radiative zone, at $r \geq 0.1814 \Rstar$, are high amplitude g modes. The dark knots observed in the bright ridges are the radial nodes of the considered mode. The number of nodes for a given mode defines the radial order $n$ of the mode, and it increases as the frequency decreases. This is an important characteristic of g modes \citep{Aerts2010}.  

The properties displayed in Fig. \ref{fig:Ekin_ref_ell20} for model \textit{ref} are common to the three boosted simulations. We find that IGWs are dampened by radiative effects and more importantly that the damping strength depends on the location in radius, as can be expected from the radiative diffusivity profile showed in Fig. \ref{fig:Krad_profile}. The damping rate seems relatively constant between the convective boundary and $r \simeq 0.8 \Rstar$, but above this radius, wave damping appears to be strengthened. This means that waves deposit their energy in the radiative cavity but there are regions where this deposition will be more important, particularly close to the top of the simulation domain. This will be further investigated in Sect. \ref{sec:wave_heating}.

\section{Waves energy flux}
\label{sec:waves_energy_flux}

As in \citet{LeSaux2022}, we define the flux $\Fsingle$ for an individual IGW mode ($\ell$, $\omega$) at radius $r$ by
\begin{equation}
\Fsingle(r,\ell,\omega) \sim \frac{1}{2}\rho \frac{N}{k_{\rm h}} P[\hat{\vel}_r^2](r,\ell,\omega).
    \label{eq:wave_flux}
\end{equation}
Then we express it in a differential form in order to compare with theoretical predictions \citep[see][for details]{LeSaux2022}:
\begin{equation}
\frac{{\rm d} \Fsingle}{{\rm d} \ln \omega {\rm d} \ln k_{\rm h}}
\sim
\frac{1}{2}\rho T_{\rm s} r N  \omega P[\hat{\vel}_r^2],
\label{eq:flux_MUSIC}
\end{equation}
with $T_s$ the sampling time.

\begin{figure*}
\centering
   \includegraphics[width=0.49\textwidth]{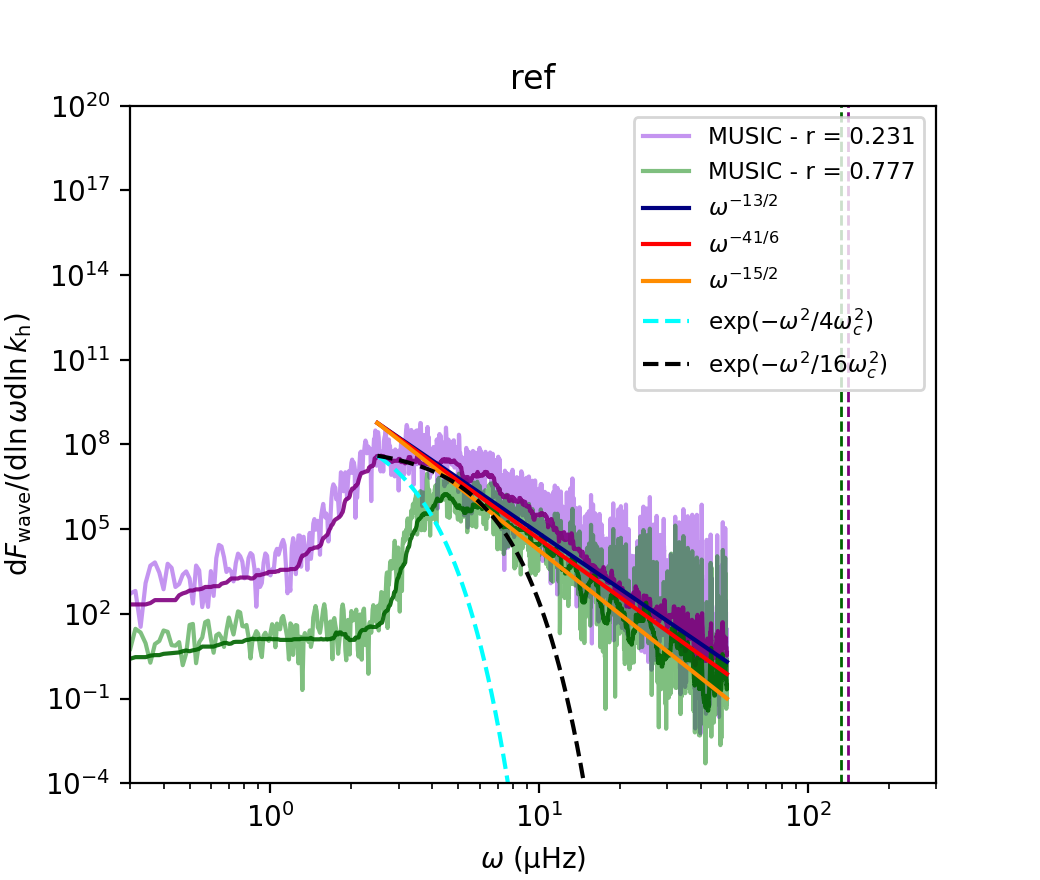}
   \includegraphics[width=0.49\textwidth]{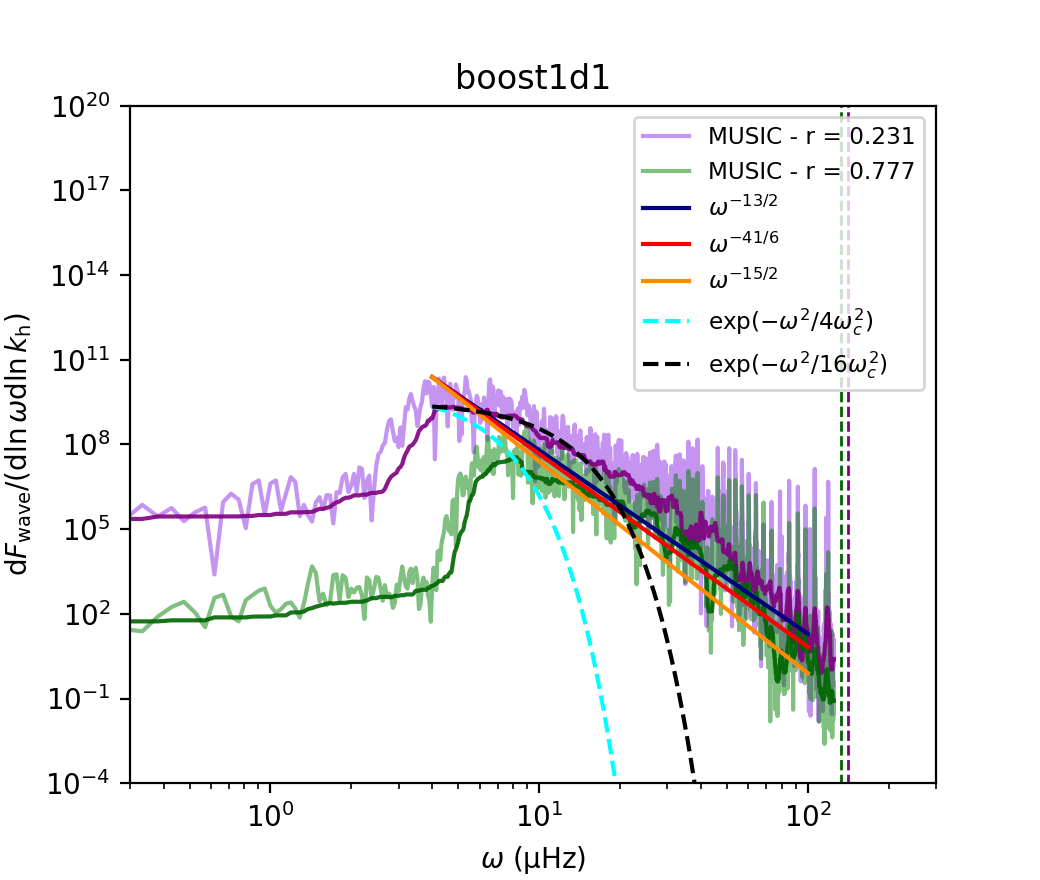}
    \includegraphics[width=0.49\textwidth]{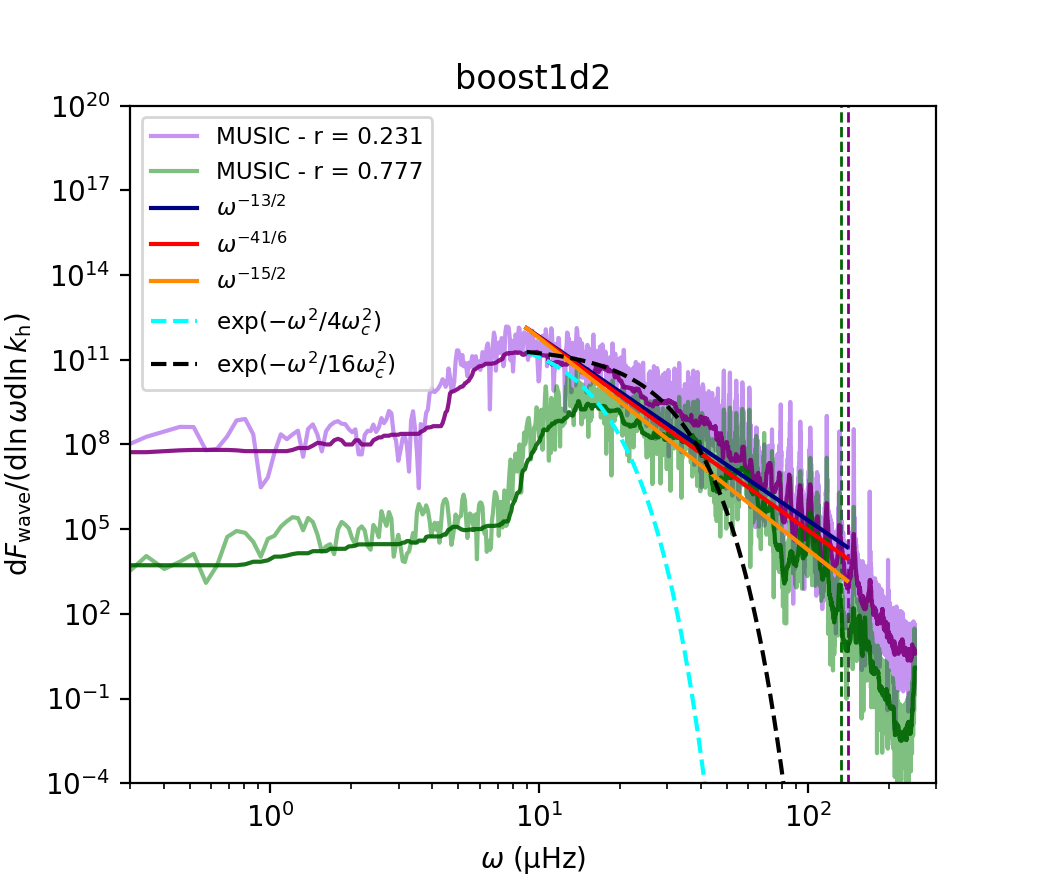}
    \includegraphics[width=0.49\textwidth]{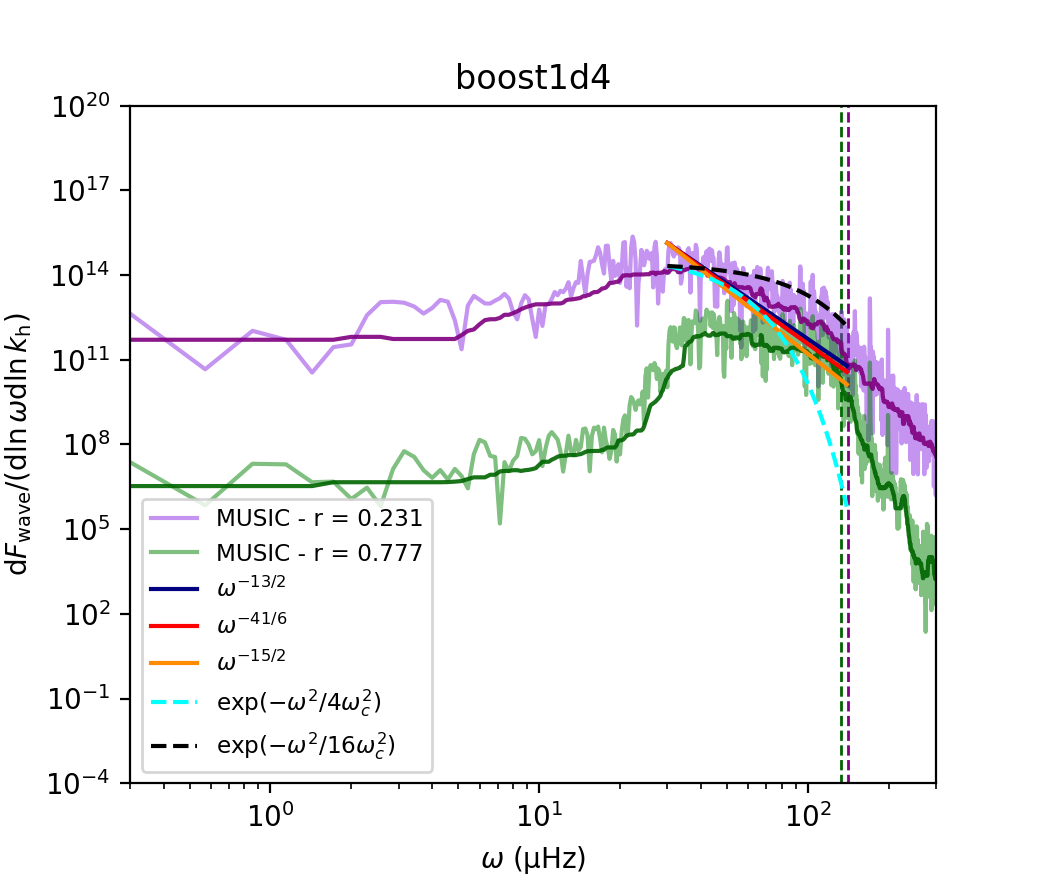}
     \caption{Wave energy flux as a function of frequency for an angular degree $\ell = 10$ for the four simulations \textit{ref} (top left), \textit{boost1d1} (top right), \textit{boost1d2} (bottom left), and \textit{boost1d4} (bottom right) at two radii, $r = r_{\rm CB} + 0.5 H_{p, \rm{conv}} \simeq 0.231 \Rstar$ (purple curves) and $r = r_{\rm CB} + 6 H_{p, \rm{conv}} \simeq 0.777 \Rstar$ (green curves). The vertical dashed lines indicate the values of the Brunt-Väisälä at $r \simeq 0.231 \Rstar$ (purple curves) and $r \simeq 0.777 \Rstar$ (green curves). The spectra are obtained via mode projection on the spherical harmonics basis and a temporal Fourier transform of the radial velocity.}
     \label{fig:powerVSfreq_theory}
\end{figure*}

Figure \ref{fig:powerVSfreq_theory} presents the dependence of the differential wave energy flux given by Eq. \eqref{eq:flux_MUSIC} with frequency at two radii, $r = r_{\rm conv} + 0.5 H_{p, \rm{conv}} \simeq 0.231 \Rstar$ (purple curves) and $r = r_{\rm conv} + 6 H_{p, \rm{conv}} \simeq 0.777 \Rstar$ (green curves). The two vertical dashed lines indicate the Brunt-Väisälä frequencies at the radii with the same colour code. The wave fluxes are plotted for an angular degree $\ell = 10$. Note that for each simulation Fig. \ref{fig:powerVSfreq_theory} shows a different frequency range as we focus on the range that bears most of the energy. The spectra present a relatively flat structure at low frequencies, then peak around a given frequency $\omega_{\rm peak}$ and finally decrease towards higher frequencies. As done in \citet{LeSaux2022}, these fluxes are compared with theoretical predictions for waves generation by Reynolds stress (blue, red and orange straight lines) and penetrative convection (cyan and black dashed curves). The analytical model used for Reynolds stress excitation is from \citet{Lecoanet2013} and for penetrative convection, or plumes, excitation the model from \citet{Pincon2016} is used. \citet{Lecoanet2013} predict the impact of the transition between the radiative and the convective regions on the wave flux. The dependence on frequency of the wave flux is modified when considering a different profile for the temperature gradient at the interface. In the case of discontinuous transition, the wave flux scales as $\omega^{-13/2}$ (dark blue solid line). If the transition is abrupt but continuous (piecewise linear) the wave flux scales with $\omega^{-41/6}$ (red solid line). Finally, if the transition is smooth (tanh profile), the wave flux scales as $\omega^{-15/2}$ (orange solid line). \citet{Pincon2016} predict a Gaussian energy flux spectrum that scales as $\rm e^{-\omega^2/4\nu_{\rm p}^2}$, with $\nu_{\rm p}$ the characteristic frequency associated with the plumes' lifetime. This frequency $\nu_{\rm p}$ is difficult to estimate analytically or from simulations, but \citet{Pincon2016} suggest that $\nu_{\rm p} \sim \omega_{\rm conv}$ (cyan dashed curve) is a good approximation. In our comparison in Fig. \ref{fig:powerVSfreq_theory} we consider a second value for  characteristic frequency associated with the plumes' lifetime $\nu_{\rm p} = 2\omega_{\rm conv}$ (black dashed curve) in order to test the influence of this parameter.
For both models, the excitation mechanism is expected to generate waves with frequencies larger than the convective frequencies, $\omega \geq \omega_{\rm conv}$. This is why we compare the wave flux from MUSIC to these analytical models in the range of frequencies $\omega \geq \omega_{\rm peak}$. For a better visual comparison with theoretical predictions, we also compute and plot a running median on the $25^{\rm th}$ percentile of the fluxes with a window of 100 frequency bins (dark purple and green lines superimposed on the corresponding flux).

After IGWs are excited at the boundary by convection in the core at frequencies $\omega \geq \omega_{\rm conv}$, they propagate away towards the surface. During this propagation, low frequency waves are damped much more rapidly than their higher-frequency counterparts. Indeed, as explained in Sect. \ref{sec:evol_ampl} radiative damping of IGWs is modelled by Eq. \eqref{eq:radiative_damping} which scales as $\omega^{-4}$. At a given radius $r$, IGWs with frequencies in the range $[\omega_{\rm conv}, \omega_{\rm peak}(r)]$ are already damped. Consequently, the peak of the flux is shifted towards higher frequencies at larger radii. This is illustrated in Fig. \ref{fig:powerVSfreq_theory}, when comparing the fluxes at two locations. For model \textit{ref} the peak is located at $\omega_{\rm peak} \sim 2.5$ \uHz{} at $r = 0.231 \Rstar$ (purple curve) and at $\omega_{\rm peak} \sim 5$ \uHz{} at $r = 0.777 \Rstar$ (green curve). Waves with frequencies between 2.5 and 5 \uHz{} have been damped before being able to reach $r = 0.777 \Rstar$. For models \textit{boost1d1}, \textit{boost1d2}  and \textit{boost1d4} the shift is from 4 to 7 \uHz{}, 7 to 10 \uHz{} and 11 to 13 \uHz{} respectively.

For frequencies larger than $\omega_{\rm peak}$, the fluxes of the four simulations are decreasing towards higher frequencies up to the Brunt-Väisälä frequency. 
Because of the high amplitude narrow peaks, which are g modes, it is not possible to precisely measure the slope of the spectra. However, for models \textit{ref}, \textit{boost1d1} and \textit{boost1d2} the wave flux measured in MUSIC is broadly consistent with analytical prediction for Reynolds stress excitation from \citet{Lecoanet2013} and for plumes excitation from \citet{Pincon2016}, but in different frequency ranges. Close to the peak the excitation seems to be dominated by penetrative convection, whereas the Reynolds stress takes over at larger frequencies. This result was already suggested by \citet{Pincon2016} (see their Fig. 3). In addition, it seems that the fit is better when using $\nu_{\rm p} = 2\omega_{\rm conv}$. However, it is difficult to disentangle the two mechanisms since Reynolds stress and penetrative convection act simultaneously to excite waves in the same frequency range, namely between $\omega_{\rm conv}$ and $N$.

Model \textit{boost1d4} shows a different behaviour. The MUSIC flux has a Gaussian shape that is broadly consistent with the plumes' excitation, but the proximity of the peak of the flux to the Brunt-Väisälä frequency makes the comparison difficult. Nevertheless, it seems that for this simulation the excitation by penetrative convection is more efficient on a larger frequency range. The shape of the curve is no longer consistent with any of the three predictions from \citet{Lecoanet2013} in any frequency range. This suggests that the excitation of IGWs  by penetrative flows is strengthened when the luminosity is increased, and seems to dominate over Reynolds stress excitation in the most boosted simulation. Interestingly, this shape of spectrum is similar to results from other multidimensional simulations, such as the ones from \citet{Rogers2013} or \citet{Edelmann2019}. In both these studies, it is suggested that the excitations of IGWs is dominated by penetrative convection. Note that the authors artificially increase the luminosity of their models by factors up to $10^7$. This suggests that enhancing the luminosity by large factors tend to increase the efficiency of plume excitation. \\

Finally, for the three models \textit{ref},  \textit{boost1d1}  and \textit{boost1d2} the slope of the wave flux remains similar close to the convective boundary (purple curve) and at the top of the domain (green curve). However, it is difficult to draw conclusions about the slope up to the stellar surface because of the wave damping that is significantly increasing from $r \simeq 0.8 R_{\rm tot}$ as shown in Fig. \ref{fig:Ekin_ref_ell20} and also because of the complexity of the near surface layers that will affect the waves. 
Moreover, when analysing hydrodynamical simulations with different values for the luminosity enhancement factor, waves of different frequencies would be able to reach the stellar surface. As we have seen earlier, higher frequencies waves will be excited with a larger amplitude and low frequencies waves are more strongly damped, when a larger enhancement factor is used. Comparing results from luminosity enhanced models with observations requires caution, since the former predict the wrong range of frequencies for waves that could reach the stellar surface. Extrapolation of spectra measured in the interior of the stellar model to the surface of the star may not be straightforward due to the strong impact of the near surface layers on waves propagation. Indeed, there is a very important increase of the radiative diffusivity and the Brunt-Väisälä frequency in this region. Consequently, we suggest that quantitative direct comparison between observations and simulations would require to run simulations as close as possible to the stellar surface. Moreover, observations of stellar oscillations do not resolve the surface of stars and consequently the signal is averaged in a way that makes it difficult to compare to simulations.

In our work we use radial velocity spectra. \citet{Lecoanet2021} suggest that it should be equivalent to use any local wave perturbation variable. We have checked this suggestion by computing spectra using temperature perturbations and co-latitudinal (horizontal) velocity and find a very good agreement with the ones computed using the radial velocity. From an observational perspective, it can be interesting to estimate the ratio of horizontal to radial velocities. At a location $r=0.85 \Rstar$ in model \textit{ref}, we calculate $\vel_{\rm h}/\vel_r \sim 270$ for $\omega =$ \uHz{10} and $\vel_{\rm h}/\vel_r \sim 13$ for $\omega =$ \uHz{40}. This decrease with frequency is expected from the dispersion relation of IGWs, which predicts that the ratio $\vel_{\rm h}/\vel_r$ varies as $N/\omega$, with the Brunt-Väisälä frequency $N$ which is fixed at a given radius. The values we obtain for the ratio of the velocities are in agreement with the ones determined in the simulations of \citet{Horst2020} in the same frequency range.
At lower frequency, the value of the ratio keeps increasing, and for $\omega =$ \uHz{2} we obtain $\vel_{\rm h}/\vel_r \sim 7800$. This value is larger than the ones calculated by \citet{Horst2020} at similar frequency. We suspect that this discrepancy is the result of the lack of independent data points from their simulations for the temporal Fourier transform, as suggested by the authors. Such high value of  $\vel_{\rm h}/\vel_r$ is in agreement with the value calculated from the two-dimensional simulations of \citet{Aerts2015}. In their simulations of rotating intermediate-mass stars, \citet{Aerts2015} obtain ratio up to $10^4$ at $r=0.99\Rstar$. 
In the observational community, this ratio is known as the K value which is approximated by $K \simeq GM/4\pi^2 \nu^2 R^3$, with $\nu$ the observed intrinsic frequency of a given star and $G$ is the gravitational constant. In their study, \citet{DeCat2002} measure this K value for SPB stars, which are mid-B type stars pulsating in high-order g modes. They obtain typical values between $\sim$10 and $\sim$100. However, note that this approximated K value is only defined at the stellar surface. Consequently, direct comparison with simulated velocity amplitudes determined deeper in the stellar interior should be taken very cautiously.

In their study, \citet{Bowman2019} are using luminosity perturbations, which is equivalent to look at the perturbations of the effective temperature $T_{\rm eff}^4$ as we have the relation $L \propto T_{\rm eff}^4$. 
First, in both spectra, observed and modelled, we note the presence of g modes, appearing as high amplitude narrow peaks. In the observed spectra, these peaks are present at frequencies larger than the so-called low-frequency power excess. We suggest that if this low-frequency power excess results from IGWs excited by core convection, we should expect to see g modes (i.e. narrow peaks in the spectra) in this low-frequency range. Indeed, in the spectrum of model \textit{ref}, there are g modes starting to appear from \uHz{10}. As we will see in Sect. \ref{sec:non-lin}, in this simulation, waves with frequencies lower than \uHz{10} are damped before being able to reach the stellar surface. However, as suggested by \citet{Edelmann2019} and \citet{Horst2020} it is also possible that the small frequency spacing between modes of different radial order and different angular degree "hides" these individual narrow peaks. To confirm this suggestion would require simulations with a radial domain extending to layers close to the stellar surface. Note that the radial extent in \citet{Edelmann2019} and \citet{Horst2020} is at 90\% and 91\% of the stellar radius, respectively.
Second, the simulation spectrum shape peaks around $\omega_{\rm peak}$ and decreases towards lower frequencies. This feature is not observed, and the spectra inferred from observations remains mostly flat at low frequencies. This difference was already reported by \citet{Edelmann2019} and \citet{Lecoanet2021}. In their study, \citet{Horst2020} state that this drop towards low frequencies is not present in the spectra measured in their simulations. They suggest that this could be attributed to the low viscosity and thermal diffusivity used in their simulations. However, our results show that this feature should be present even for non boosted simulations. We suggest that this drop is still present in the simulation of \citet{Horst2020} but it has a very low amplitude (see their Fig. 16). This is a consequence of boosting only the luminosity and not the radiative diffusivity, which will results in less damping in the frequency range close to $\omega_{\rm peak}$. 
Concerning the observed spectra from \citet{Bowman2019, Bowman2020}, this drop at low frequency is not observed. This could be the result of rotational effects that could shift wave frequency to lower values. In this case, rotation could help reconcile our simulations and observations. \citet{Rogers2013} indeed suggest an important impact of differential rotation for frequencies below \uHz{10}, based on simulations which initially impose some differential rotation. Further work including rotation and an appropriate modelling of the radiative damping of waves as close as possible  to the surface layers would be required to  confirm this effect of rotation.


\section{Non-linear effects of IGWs}
\label{sec:non-lin}

\begin{figure*}
\centering
   \includegraphics[width=0.37\textwidth]{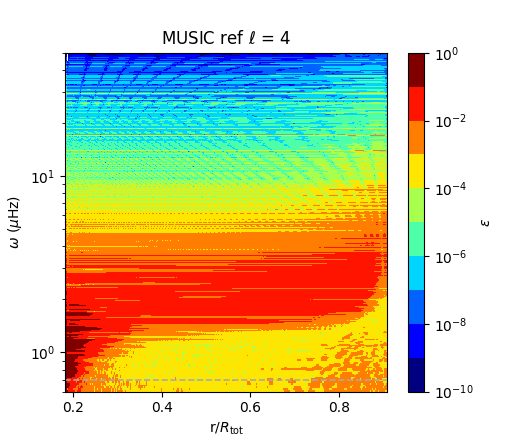}
   \includegraphics[width=0.37\textwidth]{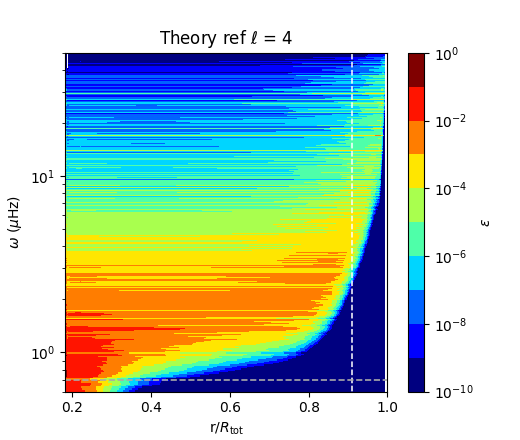}
   \includegraphics[width=0.37\textwidth]{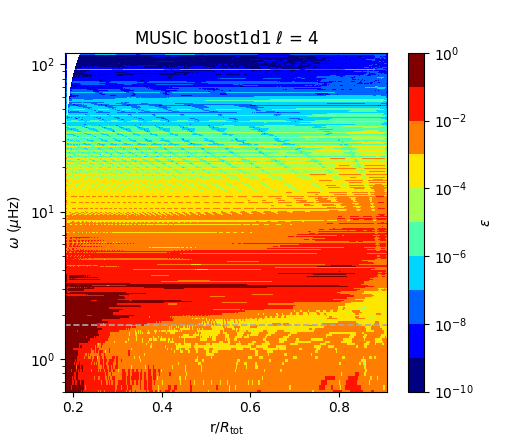}
   \includegraphics[width=0.37\textwidth]{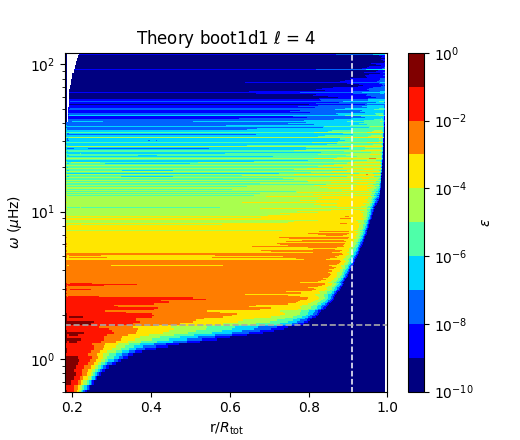}
    \includegraphics[width=0.37\textwidth]{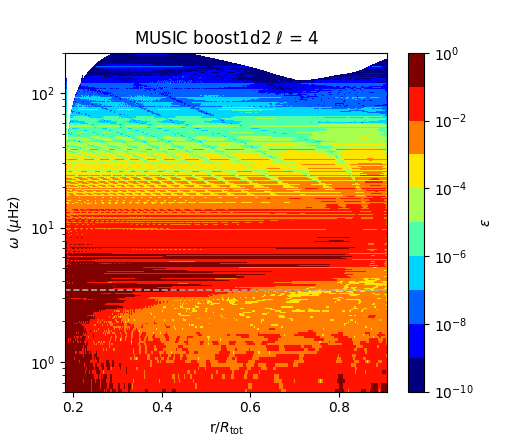}
   \includegraphics[width=0.37\textwidth]{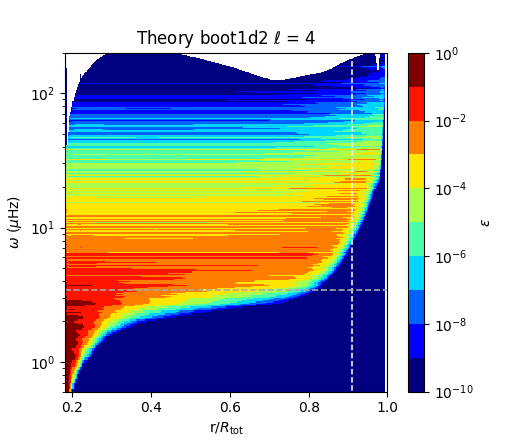}
    \includegraphics[width=0.37\textwidth]{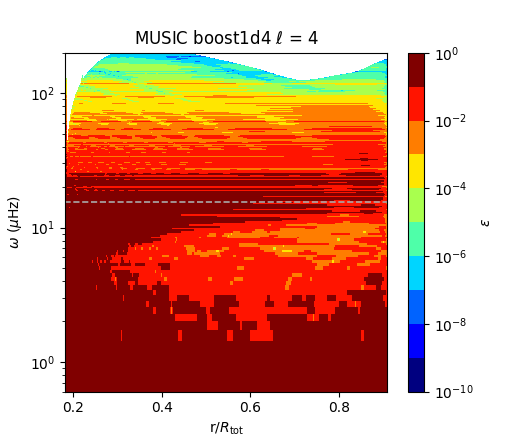}
   \includegraphics[width=0.37\textwidth]{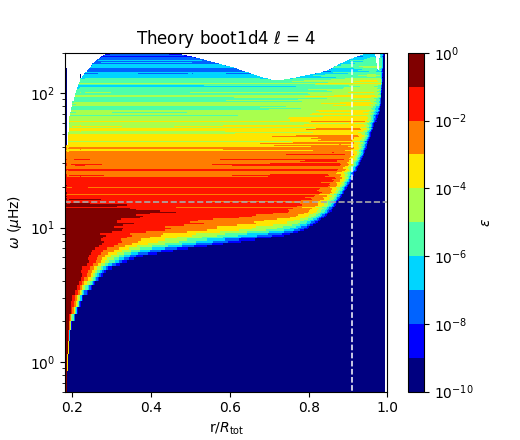}
     \caption{Dependence of the non-linear parameter $\epsilon$ on frequency and normalised radius for the four simulations \textit{ref} (top row), \textit{boost1d1} (second row), \textit{boost1d2} (third row), and \textit{boost1d4} (bottom row). The left column presents $\epsilon$ as measured in MUSIC and the right column comes from the analytical expression given by Eq. \eqref{eq:non_lin_param_theory}, normalised with the value of the velocity in the simulations at $r = r_{\rm N}$, where $r_{\rm N}$ is the smallest radius such as $\omega = N(r_N)$. The vertical white line in the analytical plots indicates the top of the numerical domain. The horizontal grey dashed line indicates the convective frequency $\omega_{\rm conv}$ for each simulation.}
     \label{fig:nonlinear_theory}
\end{figure*}

Linear theory for wave propagation is based on the assumption that the displacement amplitude of a wave is small compared to its wavelength. As the amplitude increases, non-linear effects can arise, such that the advection terms in the material derivative of the momentum equation start to play an important role \citep{Sutherland2010}. These non-linear effects can modify wave properties and evolution.
To quantify the impact of non-linear effects, we use a non-linear parameter $\epsilon$, defined as \citep{Press1981}
\begin{equation}
	\epsilon \coloneqq  \xi_{\rm h} k_{\rm h} =  \xi_r k_r,
	\label{eq:non_lin_param_displ}
\end{equation}
where $\xi_{\rm h}$ (resp. $\xi_r$) is the horizontal (resp. radial) displacement of a fluid particle associated with an IGW. This definition is basically the ratio of the horizontal (resp. radial) amplitude of a wave to its horizontal (resp. radial) wavelength. 
According to \citet{Press1981}, the displacement may be approximated as $\xi_i = \vel_i /\omega$ where $i = r, {\rm h}$. We thus have for the non-linearity parameter
\begin{equation}
	\epsilon = \frac{\vel_{\rm h}}{\omega} k_{\rm h} =  \frac{\vel_r}{\omega} k_r.
	\label{eq:non_lin_param}
\end{equation}
From this definition, an IGW is non-linear if $\epsilon \geq 1$. In the following, we compute $\epsilon$ using the definition based on the radial component of velocity and wavenumber.
Using Eq. \eqref{eq:ampl_vr}, an analytical expression for $\epsilon$ can be inferred
\begin{equation}
	\epsilon =  \vel_0 (\ell, \omega) \omega^{-1} \left(\frac{\rho}{\rho_0}\right)^{-1/2} \left(\frac{k_{\rm h}}{k_{\rm h,0}}\right)^{3/2} \left(\frac{N^2-\omega^2}{N_0^2-\omega^2}\right)^{-1/4} \rm e^{-\tau/2} k_r.
	\label{eq:non_lin_param_theory}
\end{equation}
Using the dispersion relation Eq. \eqref{eq:dispersion} for IGWs, we can calculate the radial wavenumber 
\begin{equation}
	k_r = k_{\rm h} \left( \frac{N^2}{\omega^2} -1 \right)^{1/2}.
	\label{eq:radial_k}
\end{equation}

Figure \ref{fig:nonlinear_theory} presents the dependence of this non-linear parameter $\epsilon$ on frequency and radius for an angular degree $\ell$ = 4, calculated from the MUSIC simulations (left column) and as predicted from theory (right column) using Eq. \eqref{eq:non_lin_param_theory}. We chose to analyse the angular degree $\ell$ = 4 as it is for this degree that \citet{Horst2020} predict that the non-linear effects should be most important. We have performed the same analysis for angular degrees $\ell$ = 1, 3, 5, 10 and 20, and we obtained similar results. Note that the radial range (x-axis) is different in the two columns, in the left one it extends to the top of the numerical domain $r_{\rm out} = 0.91 \Rstar$ and in the right one it extends up to the surface $r = \Rstar$. For the theoretical plots, the value of $\vel_0$ needed to normalise the amplitude of the velocity is the value of the radial velocity $\vel(r_N,\ell,\omega)$ where $r_{\rm N}$ is the smallest radius, such as $\omega = N(r_N)$ for a given frequency $\omega$. For frequencies up to $\sim 50$ \uHz{}, $r_N \simeq 0.183 \Rstar$, as in Sect. \ref{sec:evol_ampl}. For frequency between $\sim 50$ \uHz{} and the maximal value of the Brunt-Väisälä, $N_{max} \sim 220$ \uHz{}, this radius $r_N$ increases with frequency (see Fig. \ref{fig:BV_profile}). We introduce $r_N$ because of the condition for IGWs propagation $\omega < N$.
Note also that the frequency range (x-axis) available for each simulation is different. The lower frequency is set to 0.6 \uHz{} for the four simulations whereas the maximal frequencies are 50, 100, 200 and 200 \uHz{}  for \textit{ref}, \textit{boost1d1}, \textit{boost1d2} and \textit{boost1d4} respectively. This is because we  focus on the frequency range that bears most of the energy in each simulation.
We have also set a minimum threshold for $\epsilon$ at $10^{-10}$, as the actual value can become very small due to the exponential damping term. 

All plots present a similar general aspect, with horizontal ridges corresponding to waves of a given frequency and for which the value of $\epsilon$ varies with radius. 
For each simulation there is a clearly defined range of frequencies above the convective frequency (horizontal dashed grey line) with $\epsilon \geq 10^{-3}$. This occurs in different frequency ranges for each simulation. These ranges are approximately $[1.0, 5.0]$ \uHz{}, $[1.2, 9.0]$ \uHz{}, $[2.0, 20.0]$ \uHz{} and $[7.0, 50.0]$ \uHz{} for models \textit{ref}, \textit{boost1d1}, \textit{boost1d2} and \textit{boost1d4} respectively. 
For most of these waves we can see that they seem to conserve their structure in the whole envelope, suggesting that no non-linear effects occur (no energy transfer, mode coupling, etc...).

This is not the case for the lowest frequencies in the bottom region of the radiative envelope, just above the convective core. Indeed, in this region for frequencies close to the convective frequency, we can see that $\epsilon \geq 10^{-2}$, and even $\epsilon \geq 10^{-1}$ for some frequencies, both in the simulations and the theoretical plots. \citet{Ratnasingam2019} suggested that we can observe non-linear effects from $\epsilon \sim 0.1$. Therefore, we may expect non-linear effects to be relevant just above the convective core. It is possible that some IGWs generated at these frequencies have too large amplitudes and break close to the boundary.

Now, if we compare the simulations and theoretical plots, there is a major difference at very low frequencies $\omega \leq \omega_{\rm conv}$.
The simulations plots show high values of $\epsilon$ in the whole radiative zone for this frequency range, which do not appear on the theoretical plots and do not present the structure of horizontal ridges as at higher frequencies. This signal with large values of $\epsilon$ is difficult to analyse as it is localised at very low frequencies, for which the wavelengths of IGWs approach the spatial resolution of our grid. Indeed, as the radial wavelength approaches twice the length of a grid cell, it is not possible any more to represent the wave on the grid \citep[see also Sect. 6.2 of ][]{LeSaux2022}.

If we compare the four simulations, $\epsilon$ is larger as the luminosity enhancement factor is increased. This is expected as the amplitude of the waves increase with the boost. Consequently, we should expect non-linear effects to be more relevant in boosted simulations. This could result in more mixing \citep[see for example][]{Jermyn2022}, angular momentum transport \citep[see for example][]{Gervais2018} and/or wave-wave interactions. \\

Finally, by looking at the analytical predictions for $\epsilon$ up to $r = \Rstar$, we can see that non-linear effects are not expected close to the surface of the star. Most of the waves in the range of frequencies with values of $\epsilon \geq 10^{-3}$ are damped before even reaching $r = 0.91 \Rstar$, the top of the MUSIC radial domain. For waves that can propagate further, the maximal values of the non-linear parameter close to the surface are smaller than $10^{-4}$. This is consistent with the results of \citet{Ratnasingam2019} who computed a non-linearity parameter from the analytical spectra of \citet{Kumar1999} and \citet{Lecoanet2013} for Reynolds stress excitation of IGWs and found that $\epsilon < 10^{-2}$ in a 3$M_{\odot}$ star for the angular degree $\ell = 10$. However, this does not agree with the results of \citet{Horst2020} who conclude that non-linear effects may be expected at the surface of the star. We suggest that the high values of the non-linear parameter they measure in their simulation is a result of the underestimation of the radiative diffusion. Indeed, as only the luminosity of their model is enhanced, the resulting larger amplitude of IGWs is not compensated by an increase in damping.

Finally, according to the theoretical plots, some waves may reach the surface of the star, but these are relatively high frequency waves compared to the excitation frequency $\omega_{\rm conv}$. 
The predictions, however, are based on the linear theory which neglects the interaction with a possible subsurface convection zone \citep[see for example][]{Cantiello2019} and the near-surface layers that are difficult to accurately model \citep[see for example][]{Basu2017}.


\section{Wave heating in the upper layers}
\label{sec:wave_heating}

Often overlooked, thermal effects of IGWs can be significant. In the Earth atmosphere, they are known to irreversibly convert kinetic energy into internal energy \citep[see for example][]{Medvedev2003}. Through this mechanism, they can heat up the thermosphere \citep{Yigit2009}. Almost always neglected in main sequence stars, these thermal effects of acoustic and internal gravity waves have recently been accounted for in evolved stars in order to explain outbursts in supernova progenitors \citep{Fuller2017, Wu2022} and in an attempt to explain the lithium enhancement of clump stars \citep{Jermyn_Fuller2022}. In these studies, waves deposit heat through radiative diffusion.

\subsection{Temperature increase in the simulations}

\begin{figure}
    \centering
    \includegraphics[width=0.5\textwidth]{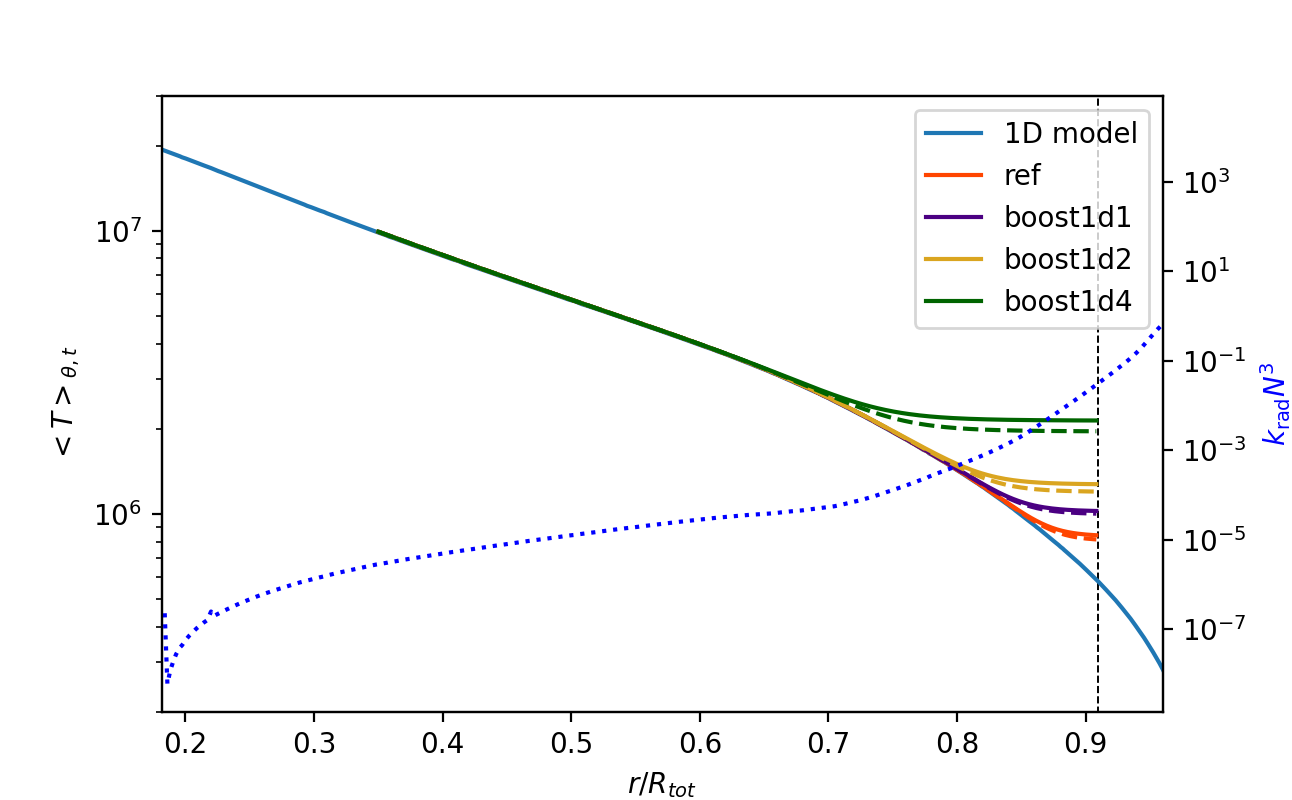}
    \caption{Spatially average temperature profile in the four simulations \textit{ref} (orange), \textit{boost1d1} (indigo), \textit{boost1d2} (yellow), \textit{boost1d4} (green). The dashed curves are the temperature profiles obtained at time $t_0$, and the plain curves at $t_0+\Delta t$, where $\Delta t$ is different for each simulation (see text for values). The temperature profile from the 1D initial model is represented by the plain blue curve. The outer boundary of the MUSIC numerical domain is indicated by the vertical dashed black line. The right axis corresponds to the term $k_{\rm rad}N^3$ (blue dotted curve) computed from the 1D model.}
    \label{fig:Heating_kradN3}
\end{figure}
In our simulations we observe a significant increase of the temperature in the upper layers of the model. Figure \ref{fig:Heating_kradN3} presents the evolution of the radial profile of the average temperature in our four simulations \textit{ref} (orange), \textit{boost1d1} (indigo), \textit{boost1d2} (yellow), \textit{boost1d4} (green). The average is performed horizontally, as defined by Eq. \eqref{eq:angular_av}.
A first average is performed at time $t_0$ (dashed curves) and a second one on $t_0 + \Delta t$ (solid curves), with $\Delta t = 7.0 \times 10^7$ s, $1.6 \times 10^7$ s, $1.2 \times 10^7$ s and $1.8 \times 10^6$ s for models \textit{ref}, \textit{boost1d1}, \textit{boost1d2} and \textit{boost1d4}  respectively.
We can clearly see an increase of the temperature close to the top of the domain in the four simulations, from $r \simeq 0.75R_{\rm tot}$ up to $r = 0.91R_{\rm tot}$.
The heated region corresponds to the one with strong wave damping observed in Fig. \ref{fig:Ekin_ref_ell20} and \ref{fig:nonlinear_theory}. The temperature increase is more important when the enhancement factor of the luminosity is larger. The term $k_{\rm rad}N^3$ (blue dotted curve) computed from the 1D model is also plotted in Fig. \ref{fig:Heating_kradN3}. In the heated region, this term increases sharply. By looking at Eq. \eqref{eq:radiative_damping} it is clear that this term drives the radiative damping of a wave at fixed $\omega$ and $\ell$. Therefore, IGWs are strongly damped in this region and will deposit a significant amount of energy. We suggest that the observed heating results from this damping by radiative diffusion.

\subsection{Theoretical estimate of heat added by waves}
In order to test this hypothesis, we analytically estimate the amount of heat $\epsilon_{\rm heat}$ added by waves in model \textit{ref} through radiative damping. Following \citet{Fuller2017}, an IGW is damped by radiative diffusion at a rate $\gamma$ given by
\begin{equation}
	\gamma \simeq k_r^2 \kappa_{\rm rad}.
	\label{eq:damping_rate}
\end{equation}
Using Eq. \eqref{eq:damping_rate}, we can define a damping length for a given wave
\begin{equation}
	l_{\rm damp} = \frac{u_{\rm{g},r}}{\gamma},
	\label{eq:damping_length}
\end{equation}
with $u_{\rm{g},r}$ the radial component of the group velocity of the wave, which can be expressed as \citep{Unno1989}
\begin{equation}
	u_{\rm{g},r} \simeq \frac{\omega^2}{N^2k_{\rm h}} \left(N^2-\omega^2\right)^{1/2}.
	\label{eq:group_vel}
\end{equation}
Using Eq. \eqref{eq:damping_rate} and \eqref{eq:group_vel} and the dispersion relation of IGWs given by Eq. \eqref{eq:dispersion}, the damping length expression becomes
\begin{equation}
	l_{\rm damp} = \frac{\omega^3}{N^2 k_{\rm h}^3\kappa_{\rm rad}} \left(\frac{N^2}{\omega^2} - 1 \right)^{-1/2}.
	\label{eq:damping_length}
\end{equation}
Then, the amount of heat deposited per unit mass per unit time by a single wave ($\ell$, $\omega$) is given by
\begin{equation}
	\epsilon_{\rm heat} = -\frac{{\rm d}L_{\rm wave}(\ell, \omega)}{ {\rm d}M} = \frac{L_{\rm wave}}{M_{\rm damp}},
	\label{eq:epsilon_heat_def}
\end{equation}
with $L_{\rm wave}$ the wave luminosity and $M_{\rm damp} = 4 \pi r^2 \rho l_{\rm damp}$ the mass through which the waves pass before being damped. The wave luminosity in Eq. \eqref{eq:epsilon_heat_def} corresponds to the luminosity initially injected by convection in a given wave, i.e. it corresponds to the initial amplitude of the wave when it is excited. Finally, radiative damping of IGWs produces a heating by unit time and unit mass for a single wave ($\ell$, $\omega$) estimated by
\begin{equation}
	\epsilon_{\rm heat}(r,\ell,\omega) = \frac{N^2 k_{\rm h}^3 \kappa_{\rm rad}}{4 \pi r^2  \rho \omega^3} \left(\frac{N^2}{\omega^2} - 1 \right)^{1/2} L_{\rm wave}(r,\ell,\omega)
	\label{eq:epsilon_heat}
\end{equation}
Using Eq. \eqref{eq:epsilon_heat} we can thus estimate an order of magnitude for the amount of heat theoretically added to the region between $r = 0.75R_{\rm tot}$ and $r = 0.91R_{\rm tot}$. To do this, we measure the wave luminosity $L_{\rm wave}$ for a mode ($\ell$, $\omega$) in our simulation \textit{ref} using
\begin{equation}
	L_{\rm wave}(r,\ell,\omega) = L_{\rm wave}(r_{\rm e},\ell,\omega) \rm e^{-\tau(r,\ell,\omega)}
\end{equation}
with $\tau$ the parameter introduced in Eq. \eqref{eq:tau} to take into account radiative damping and 
\begin{equation}
	L_{\rm wave}(r_{\rm e},\ell,\omega) = 4\pi r^2 \Fsingle(r_{\rm e},\ell,\omega)
	\label{eq:wave_lum_MUSIC}
\end{equation}
where $r_{\rm e}$ is the radius at which waves are excited and $\Fsingle$ has been defined in Eq. \eqref{eq:wave_flux}, except here we do not assume the low frequency limit. In addition, as we are interested quantitatively in the amplitude of the flux, we multiply $\Fsingle$ by a $3.28/4\pi$ to compensate for the loss of power due to FFT windowing and normalisation of the spherical harmonics. We chose $r_{\rm e} \simeq 0.183 \Rstar$ as in Sect. \ref{sec:evol_ampl} and \ref{sec:non-lin}. We consider contribution of waves with frequency $\omega$ in the range [$\omega_{\rm conv}; \omega_{\rm max}$], with $\omega_{\rm max} = 50$ \uHz{} being the maximal frequency available for the simulation. The minimal frequency is set to $\omega_{\rm conv}$ as we do not expect waves with lower frequencies to be excited \citep{Lecoanet2013}. Similarly, we consider in the estimation the contribution of waves with angular degree $\ell$ from 1 to 200.

Finally, the total amount of heat $Q_{\rm theory}$ added in the region between $r = 0.75R_{\rm tot}$ and $r = 0.91R_{\rm tot}$ during the time $\Delta t$ is
\begin{equation}
	Q_{\rm theory} = \sum_{N_{r_1,r_2}} q_{\rm th}(r),
	\label{eq:total_heat_theory}
\end{equation}
where $N_{r_1,r_2}$ is the number of radial grid cells in our simulation between $r_1 = 0.75R_{\rm tot}$ and $r_2 = 0.91R_{\rm tot}$ and $q_{\rm th}(r)$ is the amount of heat added in each radial grid cell, it is defined as
\begin{equation}
	q_{\rm th}(r) =  \sum_{\ell = 0}^{200} \sum_{\omega = \omega_{\rm conv}}^{N_{\rm max}} \rho \epsilon_{\rm heat}(r, \ell, \omega) \mathcal{V}_{\rm shell} \Delta t,
	\label{eq:radial_heat_theory}
\end{equation}
with $\mathcal{V}_{\rm shell}$ the volume of the shell between radii $r$ and $r+\Delta r$ where $\Delta r$ is the size of a numerical grid cell in the radial direction. In Eq. \eqref{eq:total_heat_theory}, $\Delta t$ is the same time interval as the one used in Fig. \ref{fig:Heating_kradN3}, which is $\Delta t = 7.0 \times 10^7$ s for model \textit{ref}. We obtain $Q_{\rm theory} = 3.3 \times 10^{41}$ erg. 

Now, in order to compare with the results from the simulations, we measure the total added heat $Q_{\rm MUSIC}$ at the top of the numerical domain using the expression
\begin{equation}
	Q_{\rm MUSIC} = \sum_{N_{r_1,r_2}} \rho c_p \Delta T \mathcal{V}_{\rm shell} 
	\label{eq:total_heat_MUSIC}
\end{equation}
with $c_p$ the specific heat capacity at constant pressure and $\Delta T$ the temperature difference after the same interval of time $\Delta t$. The total added heat in model \textit{ref} is $Q_{\rm MUSIC} = 1.0 \times 10^{43}$ erg. This value is larger than $Q_{\rm theory}$, but this can be expected. In the simulations, the top boundary $r_{\rm out} = 0.91\Rstar$ is reflecting IGWs, therefore waves that would be damped in the region between $r_{\rm out}$ and the surface in an actual star are damped in the region between $r_1 = 0.75R_{\rm tot}$ and $r_{\rm out}$ in our simulation. As previously mentioned, this heating could be strengthened due to boundary conditions. The issue of the impact of boundary conditions on hydrodynamical simulations is an open challenge that affects all stellar simulations \citep[see for example][]{Vlaykov2022}. We thus consider that the value of $Q_{\rm MUSIC}$ is in relatively good agreement with the value estimated from linear theory $Q_{\rm theory}$. This strengthens our confidence that the heating observed in the simulations results from the damping of IGWs by radiative diffusion.

\begin{figure}
    \centering
    \includegraphics[width=0.5\textwidth]{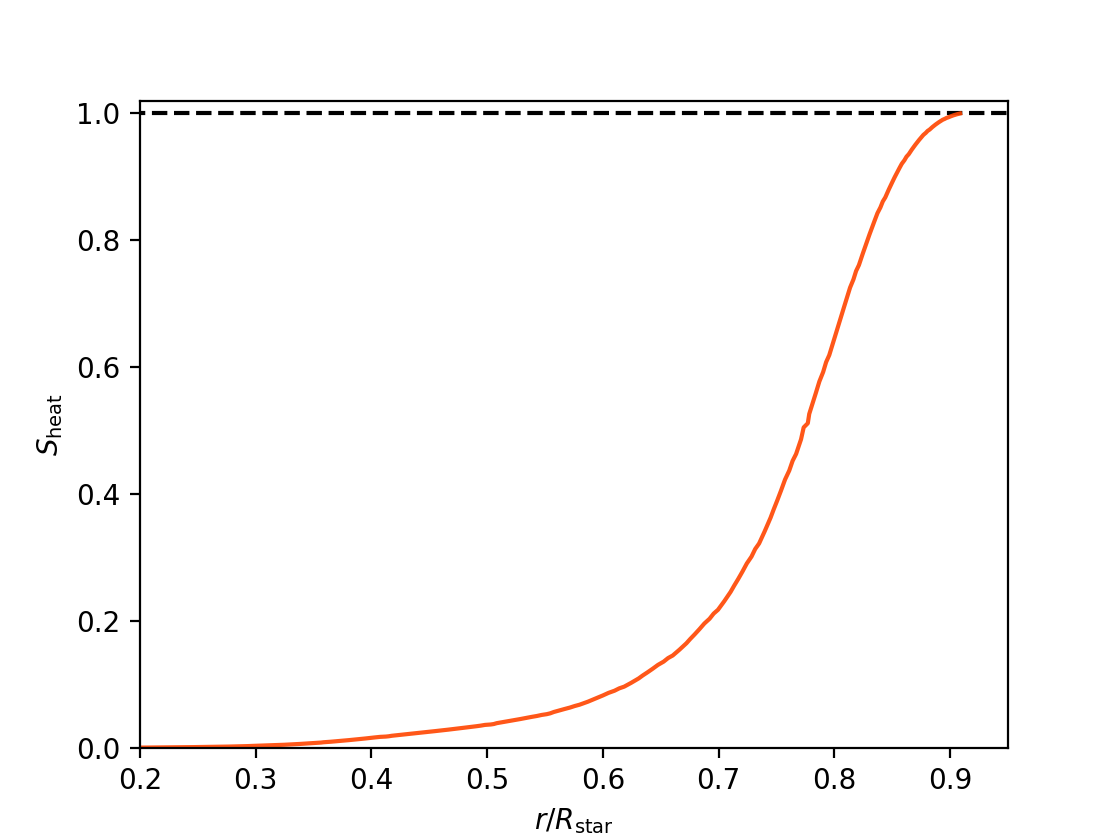}
    \caption{Cumulative sum of the theoretical estimate of the heat deposited by waves, $S_{\rm heat}$, as a function of normalised radius. The horizontal dashed line indicate the value $S_{\rm heat} = 1.0$.}
    \label{fig:CDF_heating}
\end{figure}

The estimation of $Q_{\rm theory}$ is performed between $r_1 = 0.75\Rstar$ and $r_2 = 0.91\Rstar$. However, the location of the inner boundary of this domain has a limited influence on the computation of $Q_{\rm theory}$. This is highlighted in Fig. \ref{fig:CDF_heating}.
This figure presents the radial cumulative sum of added heat by IGWs, $S_{\rm heat}$, estimated with the analytical expression given by Eq. \eqref{eq:total_heat_theory}. It is  defined as
\begin{equation}
S_{\rm heat}(r) = \frac{\sum_{N_{r_0,r}} q_{\rm th}(r)}{Q_{\rm theory}}
\end{equation}
where $N_{r_0,r}$ is the number of radial grid cells between $r_0=0.2\Rstar$ and a given radius $r$.
We can see that most of the heat that waves can deposit is at radii $r \geq 0.7 \Rstar$. This corresponds to the region where we observe an increase of the temperature in Fig. \ref{fig:Heating_kradN3} for model \textit{ref}.

In summary, our results show an increase of the temperature in the upper layers of the star. The amount of heating is in agreement with theoretical predictions of heat added by IGWs damped by radiative diffusion. This thermal impact of IGWs may thus be relevant in main sequence stars and is currently under investigation.


\section{Discussion and Conclusion}
\label{sec:discussion}

This study presents an analysis of the properties  of IGWs excited by convection in a two-dimensional fully compressible simulation of a 5 solar mass star model at zero-age-main-sequence. Our reference simulation is run with a luminosity that is not artificially enhanced and with a realistic radiative diffusivity profile. The simulation radial domain extends from $r_{\rm in}=0.02\Rstar$ to $r_{\rm out}=0.91\Rstar$.
We have highlighted that waves propagating in the radiative envelope of such stars are strongly damped by radiative diffusion. In our truncated model, the radiative diffusivity varies radially by 5 orders of magnitude. In the region not modelled in this work, between $r_{\rm out}$ and $\Rstar$, we expect the effect of radiative damping to be even more important, since the radiative diffusivity varies by more than 4 orders of magnitude in this region. Our analysis highlights the importance of including radiative diffusion in stellar hydrodynamical simulations. This is particularly important if the goal is to analyse the waves that can reach the stellar surface and to establish a link with observations or to study their transport properties. A limitation of our work is that we do not include rotation in our simulations, since the primary goal is to analyse the effect of radiative damping with realistic radiative diffusivity  profiles. A relevant comparison of simulated spectra with observations is a challenge as it would require a proper description of the surface layers and including rotation. Indeed, many OB stars observed showing photometric variability are rotating \citep[e.g.][]{Szewczuk2021, Pedersen2021}. In addition to the dynamical effects of rotation on convection (e.g. impacting $\omega_{\rm conv}$ and the plume dynamics) and thus on wave excitation, rotation may produce a shift in the wave frequencies. As already mentioned, \citet{Rogers2013} suggest a large impact of differential rotation on the low frequency power spectrum. But these conclusions are based on an imposed differential rotation  profile and would need confirmation with further numerical simulations. Interestingly, the shape and structure of the power excess observed in these stars\citep[e.g.][]{Szewczuk2021, Bowman2019} is very similar for most stars in their sample. Since the rotation rates of B dwarfs vary by more than one order of magnitude, it is compelling to observe such similarity, since the expected frequency shift would depend on the rotation rate.

In the simulations presented in this work, we observe an increase of the temperature close to the top of the numerical domain. In this region the damping of IGWs is strong, due to the simultaneous increase of stratification and radiative diffusivity. In the geophysics community, it is well known that IGWs can heat up the atmosphere. Using the linear theory of IGWs, we have estimated an order of magnitude for the amount of heat that could be added by waves in these upper layers. This value is comparable to the value inferred from the MUSIC simulations, suggesting that IGWs may be at the origin of the observed heating. Despite the fact that this heating could be slightly enhanced by the reflection of IGWs on the top boundary of our numerical domain, our results show that it is not an artefact.
Since the radiative diffusivity can vary steeply with radius, wave damping is not uniform throughout the star and waves will deposit different amounts of energy at different radii. Once again, this highlights the importance of using a realistic radiative diffusivity profile. However, this heating induced by waves is, to our knowledge, always neglected in main-sequence stars. This may not be always justified and thermal effects of IGWs may be relevant in some cases, particularly as this occurs in the outer part of the star. In a future work, we will study in which context this has to be taken into account and how it could impact stellar structures.

We have also studied non-linear effects linked to IGWs propagating in the radiative envelope of intermediate-mass stars. Waves can be efficiently excited by turbulent convection, with a large amplitude, and as they travel towards the surface IGWs see their amplitude growing as the density of the medium is decreasing. Their amplitude is also decaying due to damping by radiative diffusion, and it turns out that this effect is dominant for large amplitude IGWs, i.e. those generated with frequency close to the convective frequency. Our results suggest that non-linear effects may be relevant only above the convective core, close to where waves are excited. In this region, waves with frequencies close to the convective frequency appear to be strongly damped or to break in MUSIC simulations. More precisely, these waves have a non-linear parameter $\epsilon$ close to 1, meaning that they may be highly non-linear. A possible explanation could be that these waves are generated with a very large amplitude and that they likely break almost immediately due to non-linear effects, just above the convective core. This could result in mode coupling and/or generation of lower frequency waves. However, we cannot exclude that the signal in this very low frequency range results from aliasing as the associated wavelengths may not be resolved properly on the numerical grid of our model. This is a common issue in hydrodynamical simulations that is difficult to quantify.

The results obtained for our reference model \textit{ref} are compared with three simulations for which the luminosity has been increased by factors 10, $10^2$ and $10^4$. This comparison highlights the impact of this artefact on the generation and propagation of IGWs. In stars, the main IGWs excitation mechanism is linked to convection and has two components that are Reynolds stress and penetrative convection, both acting simultaneously. Both these mechanisms generate waves in the same frequency range. It is however difficult to determine whether one mechanism is more efficient than the other. In our simulation with realistic luminosity and radiative diffusivity, the wave flux measured is broadly consistent with excitation by penetrative convection at low frequency and with Reynolds stress at higher frequencies. This is similar in the two less boosted models but not in the most boosted one. In the latter, the wave energy flux is approximately consistent with an excitation by penetrative convection but not by Reynolds stress. This suggests that the efficiency of these two excitation mechanisms is also impacted by an enhancement of the luminosity and radiative diffusivity.
We have also shown that the heating close to the top of the domain and non-linear effects are more important when the luminosity is artificially increased. In addition, this should also impact the transport of angular momentum, which depends on the radiative damping of IGWs and on their amplitude. Results are thus quantitatively modified by an enhancement of the luminosity and any prediction based on this artefact should be taken with caution.

Concerning the surface manifestation of IGWs excited by core convection, our main conclusion is that extrapolating simulated spectra determined at an internal radius to the surface in order to compare to observations is likely meaningless. Such a comparison would require numerical simulations extending up to the surface layers, to properly describe the radiative damping in these layers, which is a formidable challenge for stellar hydrodynamics simulations.
Finally, we can identify two additional challenges regarding the comparison between observations and numerical simulations. First, simulations require physical simplifications and are thus far from realistic stellar conditions. As shown in this work, these assumptions can impact the physics of waves. Second, observations do not resolve stellar surface and thus only consider global variations of luminosity. Further efforts are thus needed in order to improve the reliability of comparisons between observations and the predictions of numerical simulations.

\section*{Acknowledgements}
We thank Jim Fuller and Adam Jermyn for interesting discussion about wave induced heating. This work is partly supported by the consolidated STFC grant ST/R000395/1 and the ERC grant No. 787361-COBOM. The authors would like to acknowledge the use of the University of Exeter High-Performance Computing (HPC) facility ISCA and of the DiRAC Data Intensive service at Leicester, operated by the University of Leicester IT Services, which forms part of the STFC DiRAC HPC Facility. The equipment was funded by BEIS capital funding via STFC capital grants ST/K000373/1 and ST/R002363/1 and STFC DiRAC Operations grant ST/R001014/1. DiRAC is part of the National e-Infrastructure. Part of this work was performed under the auspices of the U.S. Department of Energy by Lawrence Livermore National Laboratory under Contract DE-AC52-07NA27344.

\section*{Data Availability}

The data underlying this article will be shared on reasonable request to the corresponding author.



\bibliographystyle{mnras}
\bibliography{references} 






\bsp	
\label{lastpage}
\end{document}